# Resolving satellite–*in situ* mismatches in Net Primary Production using high-frequency in situ bio-optical observations in the subpolar Northwest Atlantic

Kam K.[1,*], Devred E.[3], Clay S.[3], Amirian M.[2], Irwin A.[2], Atamanchuk D.[1],Send U.[4], Wallace D.W.R[1]

*[1] Department of Oceanography, Dalhousie University, Halifax, NS, Canada*
*[2] Department of Statistic, Dalhousie University, Halifax, NS, Canada*
*[3] Ocean and Ecosystem Science Division, Fisheries and Oceans Canada, Bedford Institute of Oceanography, Dartmouth, NS, Canada*
*[4] New York City College of Technology, City University of New York, Brooklyn, NY, USA*
*[5] Scripps Institution of Oceanography, University of California San Diego, La Jolla, CA, United States*

*Corresponding author: Kitty.kam@dal.ca (Kitty Kam)

# Abstract

Net primary productivity (NPP) plays a central role in biological carbon pump (BCP) and global carbon cycling, but NPP estimates in high-latitude regions remain highly uncertain despite their disproportional contribution to the global carbon sink. Previous studies identified discrepancies among satellite and in situ-derived NPP estimates, but the drivers of disagreement remain poorly resolved due to limited high-frequency observations. Here, we present continuous depth-resolved estimates of in situ NPP in the subpolar Northwest Atlantic (56°N), reconstructed from daily bio-optical profiles collected between May and October 2016, and a 40-years dataset of photosynthesis parameters derived from 14C photosynthesis-irradiance (P-I) incubation experiments. Comparing with two satellite-based models showed satellite models overestimated NPP by factors of 2.5 – 4 from the study period. Statistical analyses identified the parameterization of the P-I relationship as the primary source of disagreement between satellite and in situ NPP estimates. Given the importance of NPP to estimates of the biological carbon pump, our findings highlight the need for regionally appropriate P-I parameterization in NPP models, particularly in high-latitude regions.

# 1 - Introduction

Marine primary productivity makes up 50% of global primary productivity and provides a baseline for estimation of the potential export and sequestration of biologically-metabolized carbon in the deep ocean (Kulk et al., 2020; Platt & Sathyendranath, 1988). Net primary productivity (NPP) describes the total amount of organic carbon produced within the euphotic zone ($Z_{eu}$), defined by 1% of surface light level, after accounting for autotrophic respiration. Organic carbon produced by phytoplankton is either remineralized, grazed by higher trophic species, or transported into the interior ocean via gravitational sinking (Volk & Hoffert, 1985), eddy-mixing (Omand et al., 2015; Resplandy et al., 2019), the mixed-layer pump (Dall'Olmo et al., 2016) and vertical migration of animals (Fahimi et al., 2025; Steinberg & Landry, 2017), forming the bases of Biological Carbon Pump (BCP) (Boyd et al., 2019). It accounts for two-thirds of the vertical gradient of dissolved inorganic carbon concentration in the ocean, while the remaining one third carbon dioxide ($CO_2$) is sequestered through the solubility pump (DeVries, 2022). Marine carbon pumps contribute significantly to regulate atmospheric $pCO_2$ at relatively low concentration, where their absence would make up to a difference of 200 ppm of atmospheric $pCO_2$ (Parekh et al., 2006), equivalent to 1560 Pg of $CO_2$ in the ocean.

Several key metrics that are commonly used for estimating BCP, such as export ratio and efficiency, rely on the estimates of NPP (Buesseler et al., 2020). Traditionally, NPP can be estimated from the photosynthetic rate of phytoplankton using incubation of seawater samples after addition of isotopic tracers during research cruises (Balch et al., 2022).The decrease in ship-time opportunities in recent years has directly impacted these ship-based productivity measurements (Brewin et al., 2023), hence remote sensing offers a complementary approach to achieving more comprehensive data coverage (in space and time).

Satellite-based approaches for NPP estimation are constrained by three factors: (1) restricted data availability due to detection constraints (e.g., cloud coverage, low light intensity), (2) the uncertainties arising from the conversion of $R_{rs}$ to bio-optical parameters, and (3) the choice of NPP models. Insufficient sunlight, due to cloud coverage and the Earth's rotation and tilt, hinder the spatial coverage of satellite retrieval, particularly in high-latitude regions (Kahru, 2017). Its limited detection at depth (only up to and integrated over first optical depth, $Z_{pd}$) also restricts the ability to resolve the vertical distribution of bio-optical parameters within the euphotic zone (~4.6 times deeper than $Z_{pd}$) and limits the opportunity to observe subsurface phenomena, such as deep chl-a maxima (Ardyna et al., 2013; Brewin et al., 2023). Hence, satellite-based NPP models often incorporate assumptions about biomass distributions and ocean conditions (such as $Z_{eu}$).

Even when retrieval of satellite data is optimal, the propagation of uncertainties through the pipeline of NPP estimation remains a significant challenge (Brewin et al., 2023; Siegel et al., 2023). Furthermore, the choice of NPP models can lead to substantial bias in the estimation of BCP fluxes up to 100% difference (Bisson et al., 2018). Various model assumptions may not always align NPP estimates with actual biogeochemical processes in the ocean (Siegel et al., 2023). Despite these limitations, most models for carbon export estimations continue to rely on satellite-derived NPP data in their export calculations, due to the scarcity of alternative data sources. Autonomous platforms equipped with *in situ* optical and biogeochemical sensors offer the potential to address gaps in observational data (Chai et al., 2020). Unlike ship-based expeditions, these platforms can offer year-round observations with higher vertical resolution, enhancing the spatiotemporal coverage of *in situ* optical observations and quality of NPP estimates. Several comparisons of NPP estimates have been conducted between satellite and *in situ* platforms, including measurements from profiling floats (Arteaga et al., 2022; Long et al., 2021; Yang et al., 2021) and gliders (Estapa et al., 2019). Such comparisons, however, are not common in the subpolar Northwest Atlantic, despite its crucial role in global carbon sequestration.

This study aimed to identify the dominant source of bias between *in situ* and satellite NPP estimates in the subpolar ocean. We present a high-frequency, depth-resolved *in situ* NPP time series in the central Labrador Sea, derived from daily bio-optical profiles collected by a moored profiler and photosynthetic variables compiled from an archive of ship-based measurements (M. M. Amirian et al., 2026). We then compared the *in situ* NPP with two sets of satellite-based NPP estimates: one from the regionally tuned BIO model (Devred et al., 2025; Platt et al., 1991) and one from the globally applied Vertical Generalized Productivity Model (VGPM) (Behrenfeld & Falkowski, 1997). The VGPM model estimates NPP from satellite-derived chlorophyll-a, PAR, and SST, whereas the BIO model is a depth- and spectrally- resolved model that uses chlorophyll-a (Chl-a) and PAR as satellite inputs, but requires the assignment of photosynthetic parameters achieved by using satellite-based Chl-a and SST.

By using depth-resolved *in situ* observations that satellites cannot directly capture, this study evaluated how the representation of bio-optical properties and photosynthetic parameters influences NPP estimates. We identified the sources of NPP differences and determined the responsible component in satellite-based NPP model that can be improved to achieve higher consistency with depth-resolved *in situ* observations. Reconciling the discrepancy between satellite and *in situ* NPP estimates is essential for obtaining more reliable NPP estimates, and for reducing uncertainties propagated to the assessment of the BCP, marine food webs and carbon cycle.

# 2 – Study Area

Located in the subpolar Northwest Atlantic (NWA), the Labrador Sea is a strong regional carbon sink, with net annual uptake estimates of ~3 mol C $m^{-3}$ $year^{-1}$ (Arruda et al., 2024; Duke et al., 2023).In winter, intense deep convection penetrates to a depth of up to ~2000 m (Lazier et al., 2002), forming deep water masses and supporting the ventilation of the interior ocean with oxygen (Atamanchuk et al., 2020; Koelling et al., 2022). Carbon and nutrients-rich water, entrained from the interior into the surface ocean during convective mixing, plays a critical role in initiating highly productive spring blooms (Balaguru et al., 2018) and is fundamental for mediating and enhancing biological carbon export through the BCP (Martz et al., 2009).

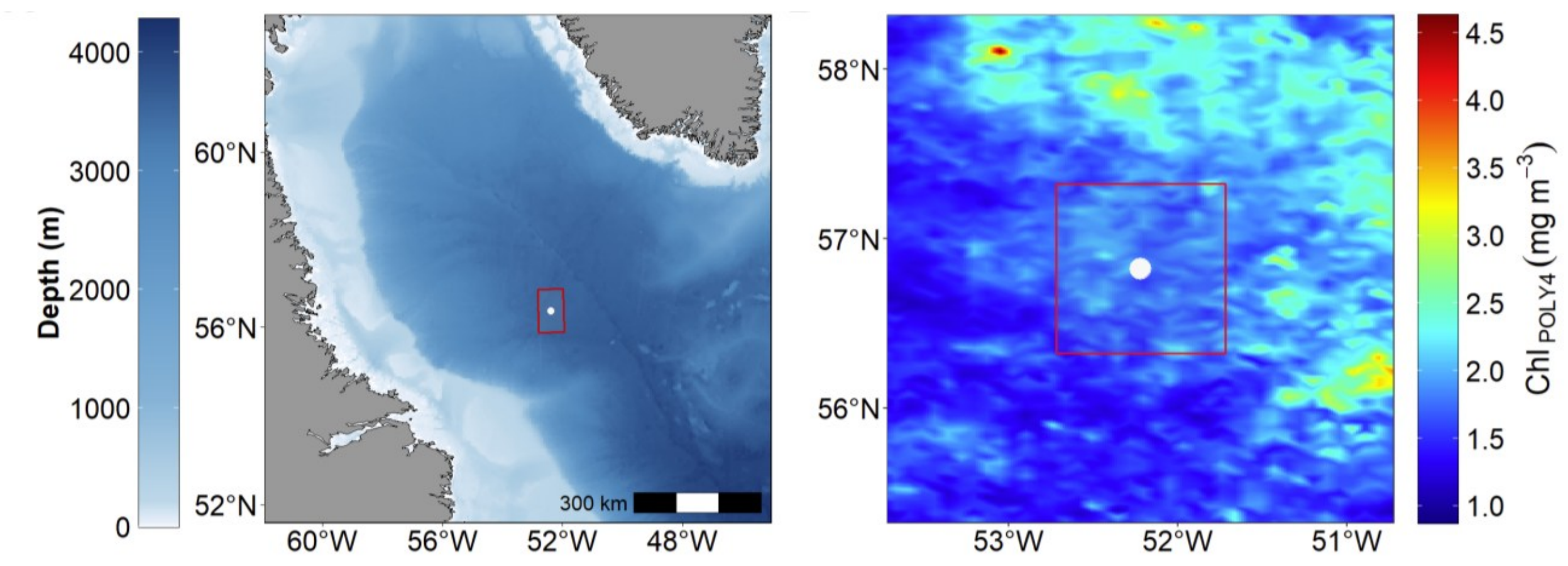

***Figure 1**. (**a**) The location of SeaCycler deployment (white dot) and the spatial coverage of remotely sensed data from MODIS-aqua satellite included in this study (red rectangle). (**b**) Satellite-based $Chl_{POLY4}$ within the observational period (May - October 2016) near the SeaCycler.*

# 3. Methods

## 3.1 Data availability

### 3.1.1. *In situ* observations: moored profiler "SeaCycler"

*In situ* profiles of bio-optical data were collected at high temporal resolution from a moored profiler, the SeaCycler, in the Central Labrador Sea (56.82° N, 52.22° W, figure 1). Profiles of bio-optical and biogeochemical parameters (Table S1), namely Chl-a concentration and photosynthetically active radiation, were measured in the upper 150 meters of the water column every 20 hours. A total of 209 bio-optical profiles (124 daytime and 85 night-time) between May and November 2016 were used to calculate *in situ* NPP.

Chl-a concentration profiles were measured from Chl-a fluorescence signals at 695 nm using a WetLabs ECO fluorescence-backscattering-CDOM (model FLBBCD2K) sensor. Chl-a fluorescence was converted to Chl-a concentration based on the linear regression of Chl-a fluorescence against extracted Chl-a (i.e., Turner Fluorometry method). Non-photochemical quenching (NPQ) was mostly observed in the top 10m of the water column due to exposure of phytoplankton biomass to high irradiance (Kiefer, 1973). Hence, NPQ was corrected for surface Chl-a by applying a combined method using night-time ratio of Chl-a to backscattering (Chl:$b_{bp}$) (Sackmann et al., 2008) and vertical extrapolation of subsurface signal to the top 10 m (Xing et al., 2012).Validation of the approach was carried out between sensor-based data and ship-based Chl-a samples collected right after the deployment of the profiler. The performance of two satellite-based Chl-a products were assessed by comparing with *in situ* Chl-a concentration ($Chl_{prof}$).

*In situ* profiles of Photosynthetically Active Radiation (PAR, integrated irradiance between 400 and 700nm) were inferred from daily profiles of downwelling irradiance ($E_d$) at 490 nm. The diffuse attenuation coefficient at 490 nm ($K_d(490)$) was calculated by regressing (Type II linear model) the log-transformed $E_d$490 against depth within the top 20m of the profiles. Finally, $E_d$ (490) was then converted to spectral $E_d(\lambda)$, with λ ranging between 400 nm and 700 nm, using a radiative transfer simulation (Jin et al., 2006). Profiles of PAR were computed by integrating $E_d$ from 400nm to 700nm at each depth.

### 3.1.2. MODIS-aqua satellite data

Remotely sensed marine bio-optical parameters were derived from Rrs measured by the Moderate Resolution Imaging Spectroradiometer onboard the Aqua satellite (MODIS-Aqua). Satellite PAR was retrieved from Top-of-Atmosphere radiances (TOA) by applying a conversion algorithm (Frouin et al., 2012) to calculate daily integrated PAR (unit: Einstein $m^{-2}$ $d^{-1}$ or mol photon $m^{-2}$ $d^{-1}$). TOA radiance was corrected for atmospheric layers to obtain Rrs, then converted to bio-optical variables using ocean-colour retrieval algorithms. Application of ocean colour algorithms to retrieve Chl-a varied with water properties. The Colour Index (CI), an empirical band difference approach, is suitable for retrieving low Chl-a concentration (<0.3 mg $m^{-3}$) (Hu et al., 2012, 2019). The Ocean Colour series (OCx), an empirical band ratio approach, is used for retrieving from high Chl-a concentration (up to 100 mg $m^{-3}$) water (O'Reilly et al., 1998; O'Reilly & Werdell, 2019).

The Chl-a product provided by NASA was processed with the combined implementation of CI and OCx algorithms (Chuanmin Hu et al., 2023). However, these approaches may contain systematic biases at the regional scale despite overall satisfactory performance at the global scale (Clay et al., 2019). This study therefore evaluated an alternative NPP output using the satellite Chl-a product processed by the "POLY4" algorithm, a regionally tuned version of the OCx model with coefficients optimized for the Northwest Atlantic (Clay et al., 2019).

## 3.2 Primary Production Models

### 3.2.1. *In situ* NPP: The Webb model

The photosynthetic-irradiance (P-I) parameters ($P^B_{max}$ , $\alpha^B$, $\beta^B$) used in this study were derived from the Amirian-tanh model (M. M. Amirian et al., 2025):

$$P^B = {P^B}_{max}\ tanh(\frac{\alpha^B\ I}{{P^B}_{max}})\ tanh((\frac{{P^B}_{max}}{\beta^B\ I})^{\gamma}) \qquad - (1)$$

where $P^B$ represents the Chl-a normalized photosynthetic rate (mg C [mg chl-a]$^{-1}$ $h^{-1}$) measured under varying irradiance levels, $I$ (W $m^{-2}$). The superscript $B$ denotes normalization to Chl-a concentration (mg chl-a $m^{-3}$), $P^B_{max}$ corresponds to the maximum photosynthetic rate (same unit as $P^B$). $\alpha^B$ represents the photosynthetic efficiency under minimal light levels (mg C [mg chl-a]$^{-1}$ $h^{-1}$ [W $m^{-2}$]$^{-1}$), $\beta^B$ represents the photoinhibition rate at high light level (same unit as $\alpha^B$) and $\gamma = \cosh^2(1) \approx 2.38$ is a fitting constant. The model was fitted to an archive of *in situ* $^{14}C$ incubation data (n=128) collected during a series of cruises across the Labrador Sea between 1977 and 2019 (M. M. Amirian et al., 2026) using the *piCurve* R package (M. Amirian, 2025; M. M. Amirian & Irwin, 2025). The photoinhibition term (B) was omitted in our study, as only 9.6% of total P-I curves exhibited minor photoinhibition, and no satellite-based approach reliably captures this effect.

*In situ* NPP was estimated using the Webb model (Webb et al., 1974) and depth-resolved Chl-a and PAR derived from the moored profiler:

$$NPP_{Webb} = \int_0^{Z_{eu}} Chl_{Prof}(z)\, P^B{}_{max}\, (1 - e^{-\frac{PAR(z) * \alpha^B}{P^B{}_{max}}})\, DL\, dz \qquad -(2)$$

where $z$ (m) is the depth, $PAR$ (W $m^{-2}$) is depth-resolved PAR profile, $Chl_{prof}$ (mg chl-a $m^{-3}$) is depth-resolved Chl-a concentration, and DL is the number of hours of daylight between sunrise and sunset computed for each day at the location of the SeaCycler (Thieurmel & Elmarhraoui, 2022). $P^B{}_{max}$ and $\alpha^B$ are assumed to be constant with depth. The depth-integrated NPP was calculated by integrating depth-resolved NPP from surface to the euphotic depth ($Z_{eu}$).

### 3.2.2. Satellite-based global model: Vertical Generalized Productivity Model (VGPM)

The Vertical Generalized Productivity Model (VGPM) is a chlorophyll-based productivity model (Behrenfeld & Falkowski, 1997) that uses satellite-derived surface Chl-a, PAR and SST to estimate column-integrated NPP using the following equations:

$$NPP_{VGPM} = Chl_{OCI} \times P^B{}_{opt} \times \frac{0.66125 \times PAR}{PAR + 4.1} \times DL \times z_{eu} \qquad -(3)$$

$$P^B{}_{opt} = \sum_{i=0}^{7} a_i \times \left(\frac{SST}{10}\right)^i \qquad -(4)$$

where optimal photosynthesis capacity, $P^B{}_{opt}$ (unit: mg C (mg Chl)$^{-1}$ $h^{-1}$), is the daily integrated production measurement as a 7-degree polynomial function of sea surface temperature (SST) based on a global relationship (Behrenfeld & Falkowski, 1997), $Chl_{OCI}$ is the NASA Chl-a product processed by the OCI algorithm (Chuanmin Hu et al., 2023) and PAR refers to the satellite-derived photosynthetic active radiation. NPP is integrated from surface to the depth of the modeled euphotic zone ($Z_{eu}$) from the Case I water model (Morel & Berthon, 1989) with the assumption of constant NPP with depth.

In this study, daily $NPP_{VGPM}$ was computed using daily measurements of satellite Chl-a, PAR and SST from MODIS-Aqua and modeled $Z_{eu}$. Satellite data was extracted and averaged within 1° radius of the SeaCycler location to maximize data availability.

### 3.2.3. Satellite-based regional model: Bedford Institute of Oceanography (BIO)

A chl-a based regional primary productivity model for the Labrador Sea (Devred et al., 2025)was updated from the Platt model (Platt et al., 1991) to derive NPP:

$$NPP_{BIO} = Chl_{POLY4} \int_0^{24} \int_0^{Z_{eu}} \frac{\Pi(z,t)}{\sqrt{1+(\frac{\Pi(z,t)}{P^B{}_{max}})^2}} \, dz \, dt \, , \qquad -(5)$$

$$\text{with } \Pi(z,t) = \int_{400}^{700} \alpha^B(\lambda) \, I(\lambda,z,t) \, d\lambda \, , \qquad -(6)$$

where *I(λ,z,t)* is the depth and wavelength-dependent light field with the subsurface light modeled using a direct and diffuse component of the irradiance in a clear sky scenario (Gregg & Carder, 1990). Hourly estimates of spectrally dependent surface light were scaled up to the daily measurement of surface PAR from the MODIS-Aqua satellite, and propagated at depth with a vertical resolution of 0.5 m. This primary production model uses satellite chl-a product derived from the POLY4 algorithm ($Chl_{POLY4}$) for the NWA region (Clay et al., 2019). Data gaps of satellite chl-a were filled using the Data Interpolating Empirical Orthogonal Functions (DINEOF) approach (Alvera-Azcárate et al., 2011).

Similar to the *in situ* NPP, regional P-I parameters ($P^B_{max}$, $\alpha^B$) were derived from fitting the same P-I Model (Eq. 1) to *in situ* $^{14}C$ incubation experiments carried out across the Labrador Sea between 2014 and 2022 (M. M. Amirian et al., 2026), using the R package "piCurve" (M. Amirian, 2025; M. M. Amirian & Irwin, 2025). Unlike the monthly averaged climatology used by the *in situ* NPP model, the spatiotemporal distribution of P-I parameters in this model was assigned by the characterization of 4 biomes based on clustering of pigments derived from High Performance Liquid Chromatography (HPLC) and delineated using satellite chl-a and SST (Devred et al., 2025). The broadband $\alpha^B$ is then multiplied by normalized phytoplankton absorption, resulting in the spectrally dependent value (Eq. 6).

## 3.3. Time Series Analysis and Model Sensitivity

This study compared NPP time series from in situ and satellite models with respect to both magnitude and seasonal variability. NPP was expressed as surface and depth-integrated values to assess how vertical extrapolation influences the magnitude of depth-integrated estimates. Surface NPP was treated as a baseline, representing the shared observational layer for both satellite and profiler data. Of the three NPP models evaluated, the in-situ Webb model was selected as the reference for NPP evaluation, although we acknowledge that this model also contains uncertainties when deriving NPP from bio-optical measurements.

Sensitivity analysis was conducted to assess the impact of model inputs (Table S2) on depth-integrated NPP by applying a systematic 10% increase on each individual input parameter and computing the percentage difference in NPP:

$$|S_{\text{NPP model}}|\ (\%) = \left| \frac{NPP_{10\%\ increase\ in\ 1\ parameter} - NPP_{Baseline}}{NPP_{Baseline}} \right| \times 100\% \qquad -(7)$$

The sensitivity of the Webb model was evaluated with *in situ* profiling data, including PAR, $Chl_{prof}$, $\alpha^B$ and $P^B_{max}$. For satellite-based models, the BIO model sensitivity was tested with satellite PAR, $Chl_{POLY4}$, $\alpha^B$ and $P^B_{max}$ ; while the VGPM model sensitivity was tested with SST, surface $Chl_{OCI}$ and PAR. Due to the limited ability of satellite optical sensors to detect subsurface features, the assumption of constant chl-a concentration in the satellite-based model was tested by replacing $Chl_{POLY4}$ to depth-resolved $Chl_{prof}$ in the BIO model and regressing the new depth-integrated output ($NPP_{BIO,\ chl\text{-}prof}$) with original $NPP_{BIO}$. This was, however, not tested in VGPM because the model lacks the P-I terms ($P^B_{max}$, $\alpha^B$) that were used in Webb and BIO.

The absolute logarithmic difference of depth-integrated NPPs between satellite-based models and the Webb model ($\Delta NPP_{VGPM\text{-}Webb}$, $\Delta NPP_{BIO\text{-}Webb}$) was assessed by applying multilinear regression analysis using the corresponding absolute differences of input variables as predictors:

$$\Delta NPP_{VGPM-Webb} = b_0 + b_1 \Delta Chl + b_2\, \Delta PAR + b_3\, P^B{}_{opt} + b_4 P^B{}_{max} + b_5 \alpha^B + b_6 t \quad -(8)$$

$$\Delta NPP_{BIO-Webb} = b_0 + b_1\, \Delta Chl + b_2\, \Delta PAR + b_3\, \Delta P^B{}_{max} + b_4 \Delta\alpha^B + b_5 t \qquad -(9)$$

Stepwise regression was conducted using Akaike information criterion (AIC) to select the best regression model by comparing models with different predictors to minimize the AIC score (Akaike, 1974). The contribution of each predictor variable in the model was assessed using a permutation test on Root Mean Squared Error (RMSE) metric, implemented in the *vip* R package (Greenwell & Boehmke, 2020). Each predictor is systematically removed from the optimized model, and the RMSE is recalculated. The data are then randomly shuffled, and a new RMSE is obtained. This process is repeated 1,000 times for each predictor, generating a distribution of RMSE values that reflects the sensitivity of model performance to the removal of that variable. Predictors that cause larger increases in RMSE are considered more influential, and their relative contributions are summarized in a Variable Importance Plot (VIP).

To investigate the impact of P-I parameters on $\Delta NPP_{BIO-Webb}$, depth-integrated $NPP_{webb}$ was recalculated with the satellite-based (i.e., SST and chl-a) assignment of P-I parameters as used in the BIO model ($NPP_{Webb^*}$ ), where the parameters were spatially and temporally matched with the profiler location and measurement time. A new time-series of depth-integrated $NPP_{Webb^*}$ was computed and compared against depth-integrated $NPP_{BIO}$. Multilinear regression analysis were applied the differences of $NPP_{Webb^*}$ and $NPP_{BIO}$ ($\Delta NPP_{BIO - Webb^*}$) using eqn. 9 but without the P-I parameters, and the contribution of the remaining variables was presented in a variable importance plot.

# 4 - Results

## 4.1. Bio-optical observations

The seasonal trend of satellite PAR is comparable to *in situ* PAR for most of the year (Figure 2a), with a moderately correlated linear relationship, except in June, early August and late September 2016 (Figure 2b, slope = 0.943). During these periods, satellite PAR was overestimated compared to *in situ* PAR, with a maximum difference of 40 mol photons $m^{-2}$ $d^{-1}$ (Figure 2b).

The seasonal trend of sea-surface Chl-a concentration shows moderate agreement between satellite-derived and *in situ* products (Figure 3a). A large phytoplankton bloom (> 5 mg $m^{-3}$), occurring between mid June and early July with a peak of Chl-a concentration at ~9 mg $m^{-3}$, was captured by both *in situ* $Chl_{prof}$ and satellite-derived $Chl_{POLY4}$. Satellite-derived $Chl_{OCI}$ captured only 50% of the concentration observed by the former two products (Figure 3a). Subsurface Chl-a structures were observed deeper than $Z_{pd}$ (Figure 3b).

Following the peak bloom in late June, spikes of Chl-a (seen as higher concentration of chl-a in bright yellow, Figure 3b) were submerged deeper than the seasonal MLD (dashed line), likely attributable to sinking phytoplankton. In mid-July, Deep chl-a maxima were detected by $Chl_{prof}$ between seasonal MLDs and $Z_{eu}$, and beyond the maximum depth detectable by the satellite ($Z_{pd}$).

*In situ* $Chl_{prof}$ exhibits a weak linear correlation with $Chl_{OCI}$ (Figure 3c, slope = 0.973, intercept = -0.837, $R^2$ = 0.47), but a stronger fit with regionally tuned $Chl_{POLY4}$ (Figure 3d, slope = 1.129, intercept = -0.569, $R^2$ = 0.54). $Chl_{POLY4}$ aligns more closely with the 1:1 line between June and August, whereas both satellite-based products displayed increasing bias from $Chl_{prof}$ starting in September and becoming most pronounced in October (Figure 3c & 3d) confirming the challenges of retrieving Chl-a at low solar angles.

## 4.2. Photosynthetic-Irradiance (P-I) Parameters

In the VGPM model, NPP output incorporates the daily optimized photosynthetic capacity ($P^B_{opt}$) based on an empirical relationship with remotely-sensed SST. As a result, the seasonal cycle of $P^B_{opt}$ directly follows the SST (Fig 4a). $P^B_{opt}$ was the lowest at ~1.8 mg C (mg chl-a)$^{-1}$ h$^{-1}$ in May 2016, and highest at ~4 mg C (mg chl-a)$^{-1}$ h$^{-1}$ in late August 2016 when SST was the highest.

In the satellite-based BIO model, the per-pixel satellite-based $P^B_{max}$ and $\alpha^B$ are higher than the monthly averages (1979-2022) used in the in situ-based Webb model, with a growing discrepancy later in the year (Fig. 4b & c). The difference in $P^B_{max}$ between the two models increased from ~2 mg C (mg chl-a)$^{-1}$ h$^{-1}$ in May 2016 to the largest difference of ~4 mg C (mg chl-a)$^{-1}$ h$^{-1}$ in October 2016 (Fig. 4b). Similarly, the difference in $\alpha^B$ increased from ~0.01 mg C (mg chl-a)$^{-1}$ h$^{-1}$ (W m$^{-2}$)$^{-1}$ in May 2016 to ~0.06 mg C (mg chl-a)$^{-1}$ h$^{-1}$ (W m$^{-2}$)$^{-1}$ in October 2016 (Fig. 4c).

Both models assume constant growth conditions vertically throughout the water column, where P-I parameters do not vary with depth. Although discrete measurements of P-I parameters showed some variability with depth, the sensitivity analysis in the Webb model (Fig. S3) suggested that the magnitude of depth-integrated NPP was not largely influenced by the depth variability of P-I parameters (up to 12.5%, Fig. S3.3) as much as the seasonality of P-I parameters (up to 250%, Fig. S3.3). For additional information, refer to the supplementary material.

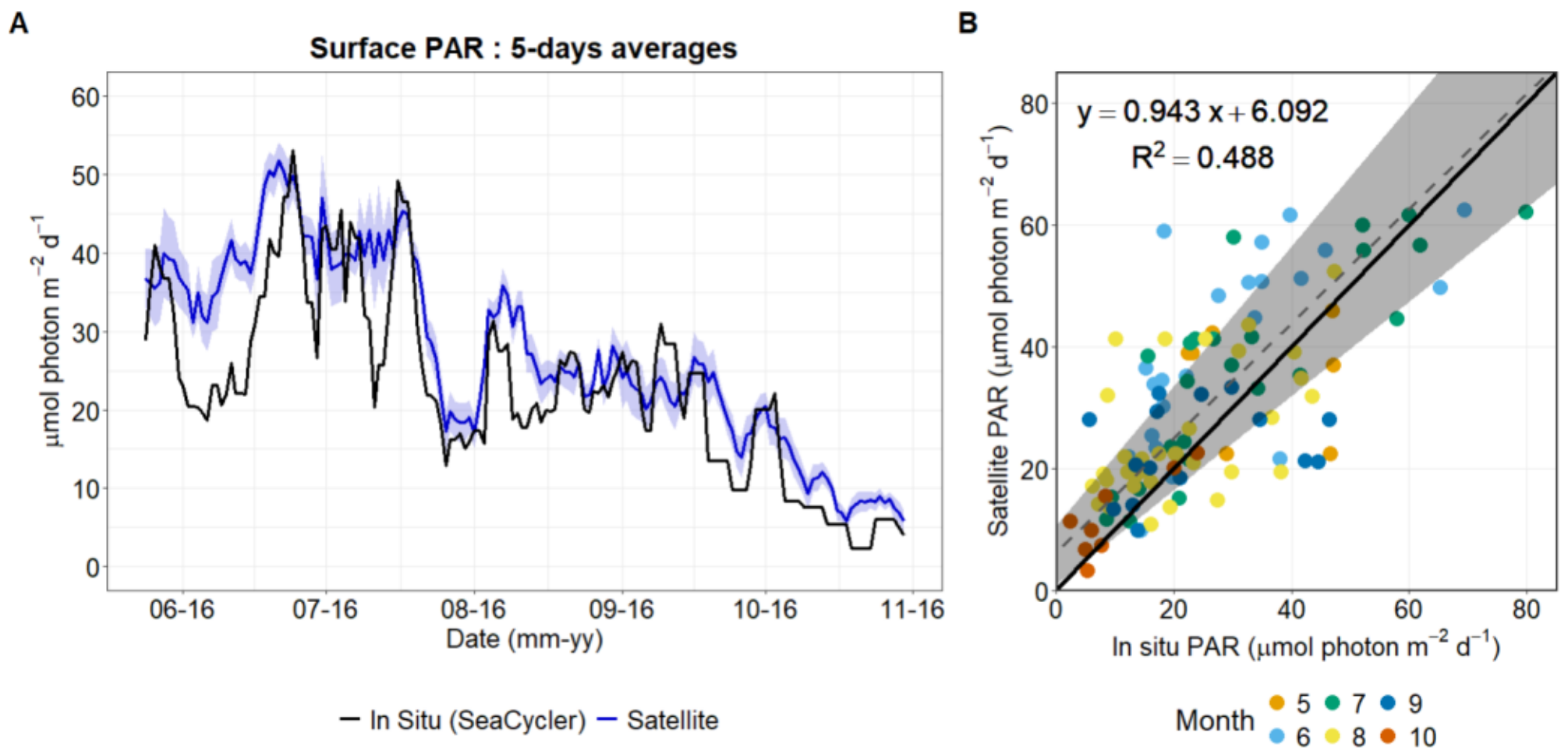

*Fig. 2. (a) 5-days averages of surface PAR from in situ (black, converted from W m$^{-2}$ to μmol photon m$^{-2}$ d$^{-1}$) and satellite-based products (blue). (b) Linear relationship between satellite and in situ PAR.*

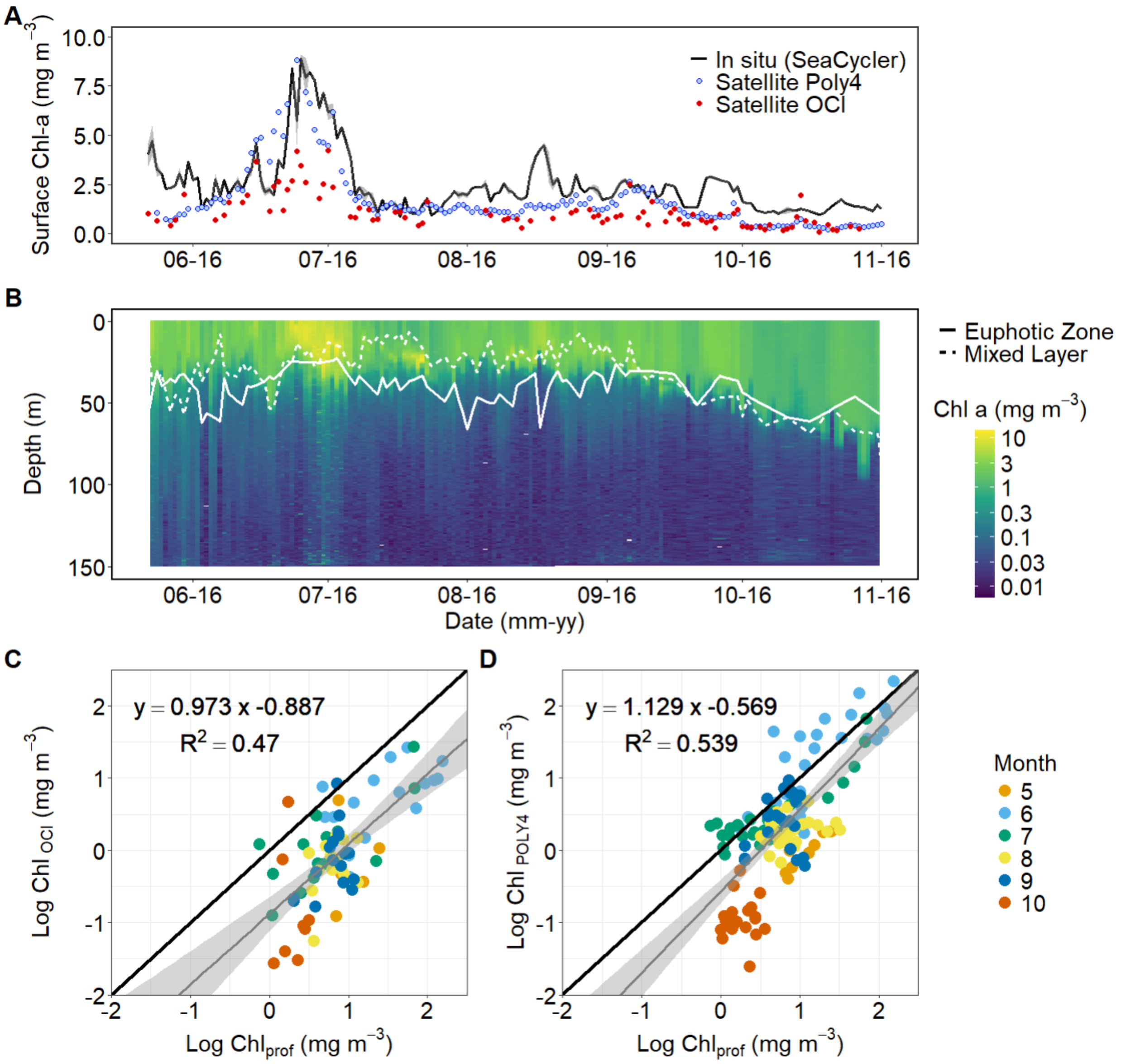

*Fig. 3. (a) Time series of surface Chl-a concentration for $Chl_{prof}$, $Chl_{OCI}$ and $Chl_{POLY4}$. Surface $Chl_{prof}$ is an average from the sea surface to the first optical depth ($Z_{pd}$ = 1/$k_{490}$), approximately 10 m on average over the seasons. (b) Time series of depth-resolved Chlprof. Solid white line is the seasonal euphotic depth (1% surface PAR) and dotted white line is the seasonal mixed layer depth (0.03 kg m$^{-3}$ deviation from surface density). Surface concentration of in situ $Chl_{prof}$ was plotted against satellite-based (c) $Chl_{OCI}$ and (d) $Chl_{POLY4}$ with colors categorized by months (May - October). The equations indicate their linear relationships with uncertainty (grey).*

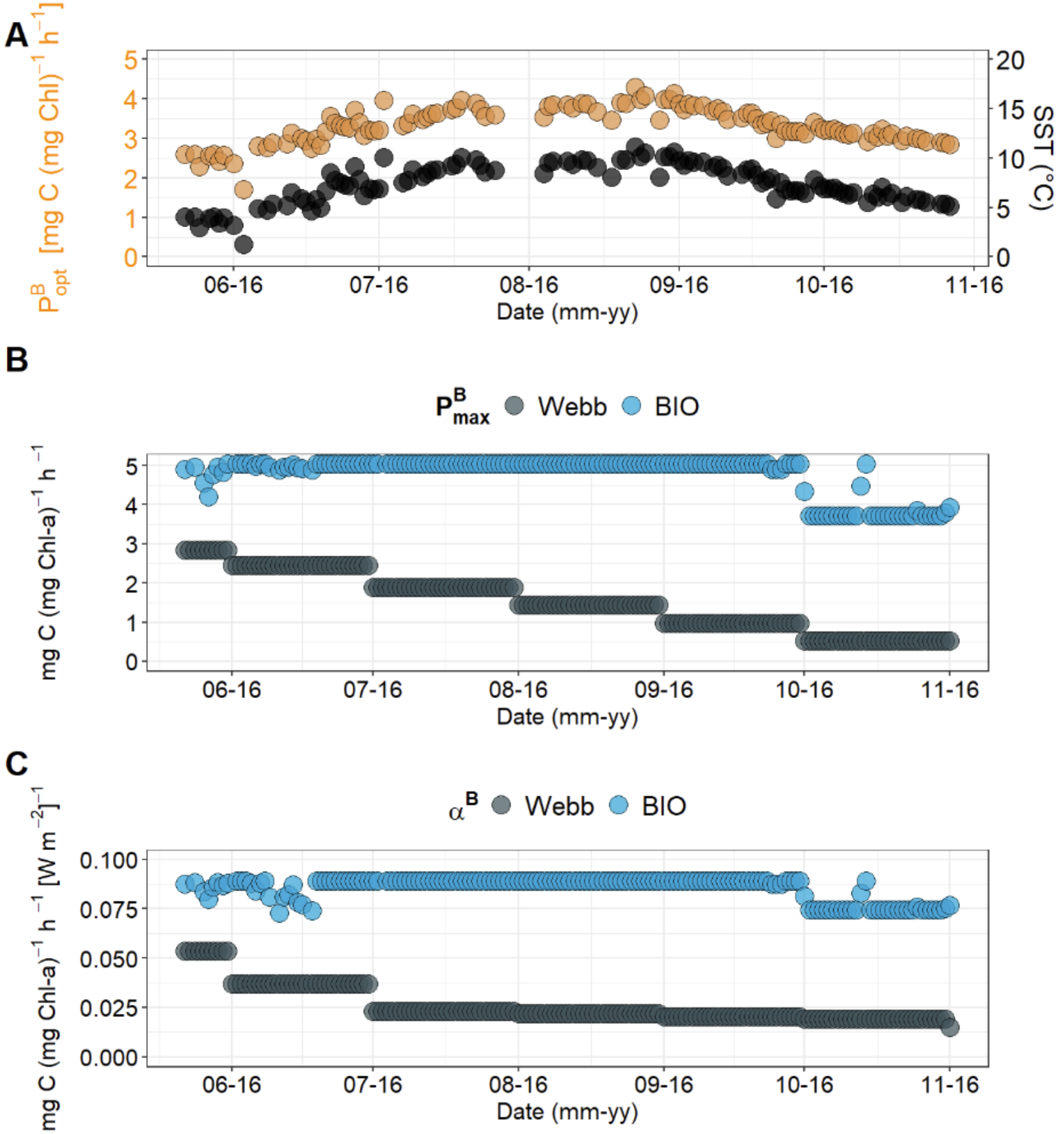

*Fig. 4. (a) Empirical daily optimized photosynthetic rate ($P^B_{opt}$) computed from satellite SST data for the VGPM model. Time series of P-I parameters: (b) $P^B_{max}$ and (c) $\alpha^B$ for in situ-based Webb model and satellite-based BIO model. The Webb model is based on monthly averages of the P-I data archives between 1979 and 2019, while the BIO model is based on spatially assigned P-I data archives between 2014 and 2022.*

## 4.3. Net Primary Production (NPP)

### 4.3.1. Comparison of NPP time series: Surface and Depth-integrated values

Surface $NPP_{BIO}$ exhibits the highest values throughout the time series (except in May 2016) (Fig 6a), followed by $NPP_{Webb}$ and $NPP_{VGPM}$. Surface $NPP_{BIO}$ reached its peak at ~55 mmol C $m^{-3}$ $d^{-1}$ in late June 2016, contributing to the largest difference of ~36 mmol C $m^{-3}$ $d^{-1}$ seen between surface $NPP_{webb}$ and $NPP_{BIO}$ (Figure 6e). In mid-September, $NPP_{BIO}$ was 4.5 times higher than $NPP_{webb}$ ($\Delta NPP_{BIO\text{-}Webb,\ Surface}$ = ~9 mmol C $m^{-3}$ $d^{-1}$). The difference between surface $NPP_{webb}$ and $NPP_{VGPM}$ ($\Delta NPP_{VGPM\text{-}Webb,\ Surface}$) remained generally within 2.5 mmol C $m^{-3}$ $d^{-1}$ except in May, and late June to early July when it reached about 5 mmol C $m^{-3}$ $d^{-1}$ (Figure 6c). This resulted from $NPP_{VGPM}$ underestimating the peak of the phytoplankton bloom, leading to a lower estimate than $NPP_{webb}$ by up to ~10 mmol C $m^{-3}$ $d^{-1}$.

Depth-integrated $NPP_{VGPM}$ and $NPP_{BIO}$ were higher than $NPP_{Webb}$ over the entire observation period (Fig 6b). The difference between depth-integrated $NPP_{VGPM}$ and $NPP_{Webb}$ (Figure 6d) was the highest in early July ($\Delta NPP_{VGPM\text{-}Webb,\ Depth\text{-}integrated}$ ≈ 200 mmol C $m^{-2}$ $d^{-1}$), contrasting with the negative difference (satellite - *in situ*) that was observed between surface NPP values (Figure 6c). The difference between depth-integrated $NPP_{BIO}$ and $NPP_{Webb}$ (Figure 6f) was highest in late June (~300 mmol C $m^{-2}$ $d^{-1}$), consistent with the timing of the maximum difference in surface NPPs between two models (Figure 6e). Despite the magnitude of $\Delta NPP_{BIO\text{-}Webb}$ being large, the seasonal trends of $\Delta NPP_{BIO\text{-}Webb}$ were similar for both surface and depth-integrated NPPs (Figs. 6e, 6f).

### 4.3.2. Model Sensitivity

The Webb model was most sensitive to $Chl_{prof}$ with $|S_{Webb}|$ equals to 10% throughout the time series (Figure 5a and b). Surface PAR and $\alpha^B$ had identical sensitivities towards $NPP_{Webb}$ (5% < $|S_{Webb}|$ <10%), resulting from the set up of the Webb model that similarly impacted both variables (Eq. 2). Surface $P^B_{max}$ was the least sensitive parameter in this model with $|S_{Webb}|$ remaining below 5%. No visible seasonal trend was observed in the sensitivity of the Webb model, where the changes of $NPP_{Webb}$ from all input variables seemed to be randomly distributed.

The BIO model was most sensitive to surface PAR (5% < $|S_{BIO}|$ <12%), followed by $\alpha^B$ (5% < $|S_{BIO}|$ <9%), $Chl_{POLY4}$ (5% < $|S_{BIO}|$ <7%) and least by $P^B_{max}$ (1% < $|S_{BIO}|$ <5%) (Figure 5c, 5d). The increase of $NPP_{BIO}$ by $Chl_{POLY4}$ was mostly stable around 5% throughout the time series. However, the BIO model seemed to be temporally sensitive to the change of PAR, $\alpha^B$ and $P^B_{max}$. After the beginning of October 2016, $NPP_{BIO}$ showed higher sensitivity to PAR and $\alpha^B$, but lower sensitivity to $P^B_{max}$ (Figure 5c). During productive seasons, characterized by high net primary productivity (NPP) values and shallower Chl-a maxima depths (<40m), the change in $NPP_{BIO}$ remained minimal as revealed by including the *in situ* Chl-a vertical profile in the BIO model (Figure S4). Pronounced shifts (up to ~50%) are observed during periods of low productivity that are associated with deeper Chl-a maxima (>40m).

The VGPM model was most sensitive to SST (4% < $|S_{VGPM}|$ <8%), followed by $Chl_{OCI}$ (5% < $|S_{VGPM}|$ <7%) and PAR (1% < $|S_{VGPM}|$ <4%) (Figure 5e, 5f). A significant seasonal trend could be seen for the sensitivity of SST in the model, with warmer temperatures resulting in higher percentage changes in $NPP_{VGPM}$ (Figure 5e). Model sensitivity of VGPM to $Chl_{OCI}$ was similar to that of the BIO model to $Chl_{POLY4}$, except for six “outliers” where $|S_{VGPM}|$ was between 1.8% and 2.2%. Moreover, model sensitivity to PAR increased after October 2016, which was also observed in the BIO model.

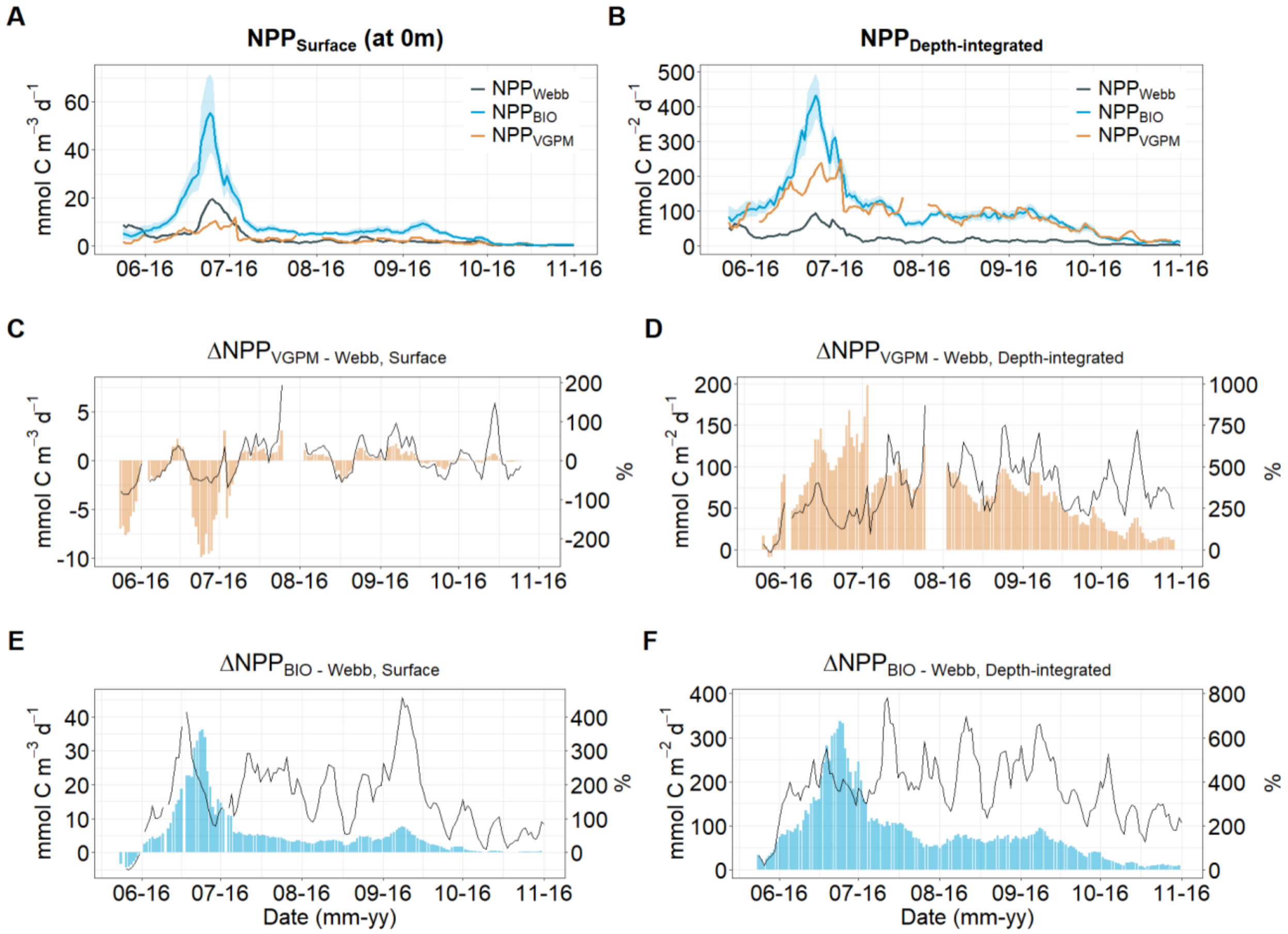

*Fig. 5. Time series in 5-days rolling averages for all (a) surface and (b) depth-integrated NPPs. Between $NPP_{VGPM}$ and $NPP_{webb}$, the differences were computed in absolute (orange bars) and percentage (black line) values for (c) surface ($\Delta NPP_{VGPM\text{-}Webb,\ Surface}$) and (d) depth-integrated ($\Delta NPP_{VGPM\text{-}Webb,\ Depth\text{-}integrated}$) time series. Similarly, the difference between $NPP_{BIO}$ and $NPP_{Webb}$*

*was computed in absolute (blue bars) and percentage (black line) values for (e) surface ($\Delta NPP_{BIO\text{-}Webb,\ Surface}$) and (f) depth-integrated ($\Delta NPP_{BIO\text{-}Webb,\ Depth\text{-}integrated}$) time series.*

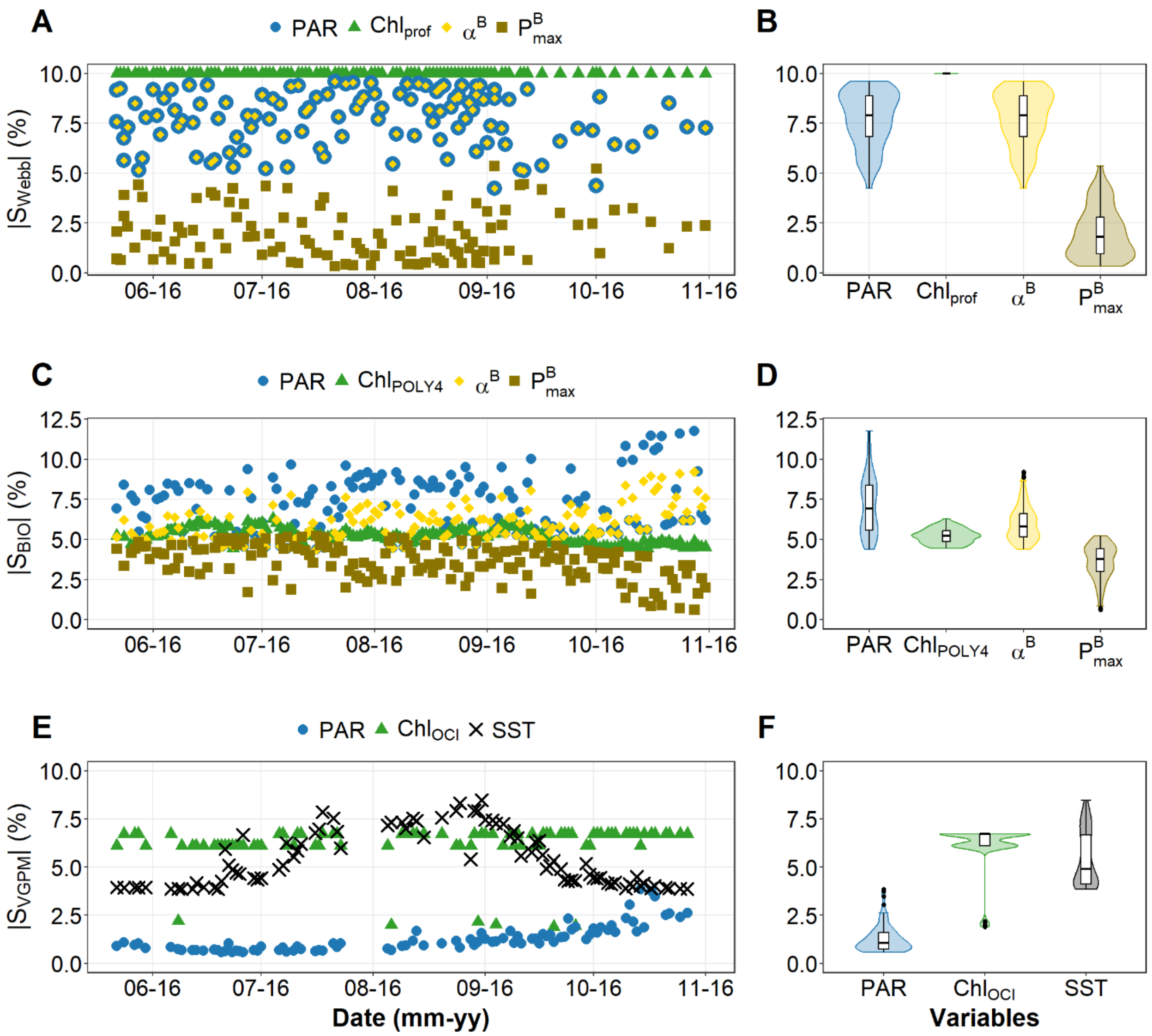

*Fig. 6. Model sensitivity (%) was assessed for (a, b) $NPP_{Webb}$ ,(c, d) $NPP_{BIO}$ and (e, f) $NPP_{VGPM}$. Temporal model sensitivities are shown in (a), (c) and (e), while the range of model sensitivity categorized by input variables are displayed in (b), (d) and (f).*

## 4.2.3. Source of NPP differences

For $\Delta NPP_{VGPM\text{-}Webb}$, $\alpha^B$ was the leading predictor with the highest importance score of 0.13 ± 0.04 followed by $P^B_{opt}$ with a score of 0.081 ± 0.031, while the importance of time, $P^B_{max}$, $\Delta Chl_{OCI\text{-}prof}$ and ΔPAR were negligible (Figure 7a). The best model selected by AIC included $P^B_{opt}$ and $\alpha^B$ as the predictors (AIC = -36.4), while all other models displayed higher AIC values (ΔAIC > 1; Table 1). The final model explained 34.5% of the variance in $\Delta NPP_{VGPM\text{-}Webb}$ ($p<0.001$; Table 2). The correlation with $\alpha^B$ was moderately negative with coefficient of b = -30.82 ± 13.28 (p-value= 0.024), whereas the correlation with $P^B_{opt}$ was marginally positive (b = 0.54 ± 0.31, p = 0.083). No clear correlation is found in the intercept of $\Delta NPP_{VGPM\text{-}Webb}$ (b = 0.43 ± 1.35, p = 0.75).

$\Delta\alpha^B$ was the most influential predictor in the model of $\Delta NPP_{BIO\text{-}Webb}$ with an importance score of 0.11 ± 0.03, followed by $\Delta Chl_{POLY4\text{-}prof}$ and time with significantly lower scores of 0.014 ± 0.010 and 0.009 ± 0.008 respectively (Figure 7b). The contributions of $\Delta P^B_{max}$ and ΔPAR were considerably less important for the model. The best regression model of $\Delta NPP_{BIO\text{-}Webb}$ only included $\Delta\alpha^B$ as the lone predictor (AIC = -69.93), as indicated by AIC selection (Table 1). However, the final model only explained 12.7% of the variance in $\Delta NPP_{BIO\text{-}Webb}$ (Adjusted $R^2$ = 0.105, p < 0.001; Table 2). $\Delta NPP_{BIO\text{-}Webb}$ was strongly correlated with $\Delta\alpha^B$ (β= 24.76 ± 6.97, p<0.001) and weakly correlated with model intercept (0.15 ± 0.42, p=0.728), indicating that the difference between $NPP_{BIO}$ and $NPP_{Webb}$ was not statistically significant.

***Table 1.*** *Regression models of $\Delta NPP_{VGPM\text{-}Webb}$ and $\Delta NPP_{BIO\text{-}Webb}$ ranked by AIC. Model in the first rank is chosen to be the final regression model with the lowest AIC score. ΔAIC is the difference of AIC score between the selected model and the final model (ranked 1st).*

| Rank | Models | AIC score | ΔAIC |
|---|---|---|---|
| | $\Delta NPP_{VGPM\text{-}Webb}$ | | |
| 1 | **$\Delta NPP \sim P^B_{opt} + \alpha^B$** | -36.4 | 0 |
| 2 | $\Delta NPP \sim time + P^B_{opt} + \alpha^B$ | -34.68 | 1.72 |
| 3 | $\Delta NPP \sim time + \Delta Chl_{OCI\text{-}prof} + P^B_{opt} + \alpha^B$ | -32.86 | 3.54 |
| 4 | $\Delta NPP \sim time + \Delta Chl_{OCI\text{-}prof} + P^B_{opt} + P^B_{max} + \alpha^B$ | -31.18 | 5.22 |
| 5 | $\Delta NPP \sim time + \Delta Chl_{OCI\text{-}prof} + \Delta PAR + P^B_{opt} + P^B_{max} + \alpha^B$ | -29.18 | 7.22 |
| | $\Delta NPP_{BIO\text{-}Webb}$ | | |
| 1 | **$\Delta NPP \sim \Delta\alpha^B$** | -58.66 | 0 |
| 2 | $\Delta NPP \sim time + \Delta\alpha^B$ | -58.33 | 0.33 |
| 3 | $\Delta NPP \sim time + \Delta Chl_{POLY4\text{-}prof} + \Delta\alpha^B$ | -57.32 | 1.34 |
| 4 | $\Delta NPP \sim time + \Delta Chl_{POLY4\text{-}prof} + \Delta P^B_{max} + \Delta\alpha^B$ | -55.33 | 3.33 |
| 5 | $\Delta NPP \sim time + \Delta Chl_{POLY4\text{-}prof} + \Delta PAR + \Delta P^B_{max} + \Delta\alpha^B$ | -53.34 | 5.32 |

***Table 2**. Statistical result of the final regression model for $\Delta NPP_{VGPM\text{-}Webb}$ and $\Delta NPP_{BIO\text{-}Webb}$.*

| Model | Adjusted $R^2$ | RSE | p-value | Predictor | Estimate (β) | p-value |
|---|---|---|---|---|---|---|
| $\Delta NPP_{VGPM\text{-}Webb}$ | 0.345 | 0.70 | **< 0.001** | Intercept | 0.43 ± 1.35 | 0.75 |
| | | | | $P^B_{opt}$ | 0.54 ± 0.31 | 0.083 |
| | | | | $\alpha^B$ | -30.82 ± 13.28 | 0.024 |
| $\Delta NPP_{BIO\text{-}Webb}$ | 0.105 | 0.74 | **< 0.001** | Intercept | 0.15 ± 0.42 | 0.728 |
| | | | | $\Delta\alpha^B$ | 24.76 ± 6.97 | **<0.001** |

.

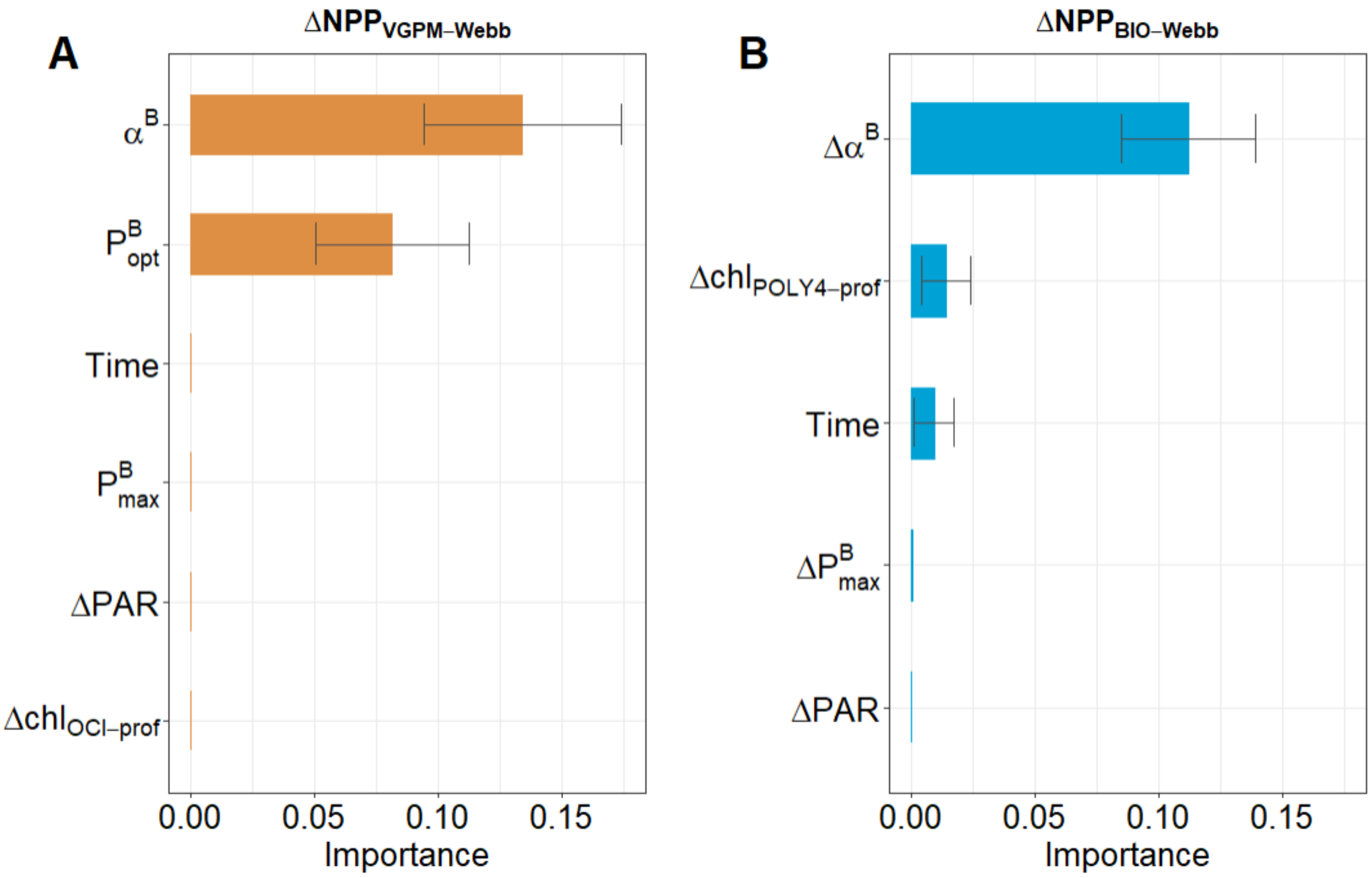

*Fig. 7. Variable Important Plot of predictors sorted by the most to least importance for the regression models (a) $\Delta NPP_{VGPM\text{-}Webb}$ and (b) $\Delta NPP_{BIO\text{-}Webb}$.*

## 4.2.4. Impact of the assignment of P-I parameters on NPP agreement

To demonstrate the impact of the P-I parameters to $\Delta NPP_{BIO\text{-}Webb}$, the regression model was recomputed with a new set of $NPP_{webb}$ which used the same P-I parameters as the BIO model ($NPP_{webb*}$). This improved the agreement between the BIO and the updated Webb models (Figure 8b). The slope of correlation increased from 0.862 (Figure 8a) to 0.998 (Figure 8b), indicating a near-perfect proportiosnal relationship between $NPP_{Webb*}$ and $NPP_{BIO}$. The intercept (on $\log_{10}$ scale) was reduced from 1.879 (Figure 8a) to 0.147 mmol C $m^{-2}$ $d^{-1}$ (Figure 8b), suggesting a significant reduction in systematic bias. Despite this, the coefficient of determination ($R^2$) decreased slightly from 0.682 (Figure 8a) to 0.603 (Figure 8b), $NPP_{webb*}$ was overall better aligned with $NPP_{BIO}$ except for some outliers in October 2016.

After AIC selection, time and $\Delta Chl_{POLY4\text{-}prof}$ were retained while ΔPAR was removed from the final model (AIC = -61.92; Table 1). Interestingly, time ranked first with an important score of 0.05 ± 0.02 as the most influential variable to the difference between $NPP_{BIO}$ and $NPP_{Webb*}$ (Figure 8c), while $\Delta Chl_{POLY4\text{-}prof}$ was the driving parameter for $\Delta NPP_{BIO\text{ - }Webb}$ (Figure 7b and Table 1). $\Delta Chl_{POLY4\text{-}prof}$ ranked second with an importance score of 0.015 ± 0.011, three times less important than time, while the impact of ΔPAR in $\Delta NPP_{BIO\text{ - }Webb*}$ was negligible.

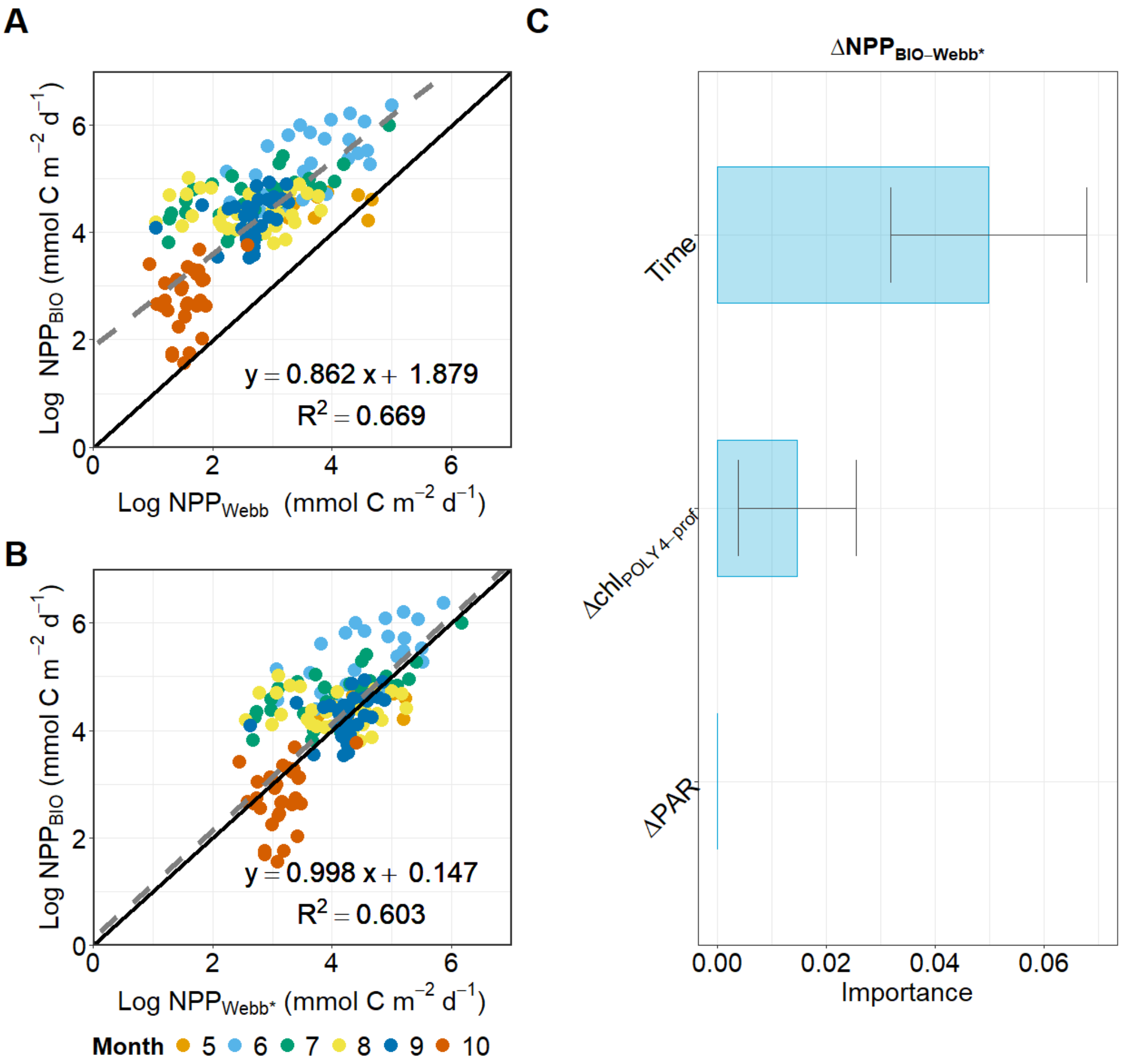

*Fig. 8. Linear relationship between $NPP_{BIO}$ and $NPP_{webb}$: (a) Depth-integrated $NPP_{webb}$ from monthly averages of P-I parameters, and (b) updated depth-integrated $NPP_{webb}$ (denoted as $NPP_{Webb*}$) using the assigned P-I parameters for $NPP_{BIO}$. (c) The Variable Importance analysis ranked by importance score for the regression model of $\Delta NPP_{BIO\text{ - }Webb*}$.*

# 5 - Discussion

Comparisons of satellite-based productivity models have been previously carried out during round-robin exercises on a global scale and in the Arctic regions (Campbell et al., 2002; Carr et al., 2006; Y. J. Lee et al., 2015; Regaudie-de-Gioux et al., 2019; Saba et al., 2010). However, there are very few comparisons performed using *in situ* NPP derived from a mooring equipped with bio-optical sensors. In this study, *in situ* $NPP_{Webb}$ was selected as the reference to compare with two satellite-derived NPP estimates because it was constrained by high-frequency, depth-resolved observations of Chl-a and PAR. The Chl-a profiles were validated with *in situ* ship-based Chl-a measurements and corrected for non-photochemical quenching, providing a better representation of temporal and vertical variability that cannot be observed by satellite. Using this comparison framework, we evaluated model sensitivity to input variables and examined the differences in these variables associated with the disagreement among NPP estimates. While several satellite-based primary production models exist, the study was limited to two models: the VGPM and the BIO model.

## 5.1. Model sensitivity - what drives the model most?

Chl-a concentration exerts the strongest influence on $NPP_{Webb}$ among other inputs, while satellite-based models are 40 to 50% less sensitive to Chl-a concentration than the Webb model. Although *in situ* $Chl_{prof}$ was better correlated with $Chl_{POLY4}$ than $Chl_{OCI}$ in the top 10m of the water column, both BIO and VGPM models have similar ranges of sensitivity to Chl-a concentration (i.e., about 6%). This is assumed to be tied to the vertical resolution of Chl-a between *in situ* and satellite products. Both BIO and VGPM models assume constant Chl-a concentrations in the water column and do not consider the subsurface structure of Chl-a. Although efforts have been carried out to derive Chl-a vertical profiles in the water column using a statistical approach (Platt et al., 1991), these were not implemented in the current study.

In other regions, such as the Arctic, the incorporation of Chl-a maxima reduced the error of satellite-based NPP estimates by 25% (Matthes et al., 2023). Our study showed that subsurface Chl-a structure can change up to ~50% at the lower end of depth-integrated $NPP_{BIO}$ that correlates with Chl-a maxima deeper than 40m (Figure S4), highlighting that the vertical structure of Chl-a profile could have a higher impact on satellite-based NPP estimates. A correction factor of 0.83 (Figure S4) on NPP outputs could be considered for satellite NPP models in this region during the seasons when deep Chl-a maxima and low productivity have been observed.

The BIO model is largely driven by PAR and $\alpha^B$, which the Webb model also shares similar sensitivity to both variables. At higher latitudes, light intensity is significantly reduced compared to the lower latitudes. As the light penetration in the euphotic zone changes seasonally, the light availability could be the limiting factor of phytoplankton growth at the high latitude, particularly in spring when nutrients in the water column are plentiful. Therefore, it would be reasonable that both models are more sensitive to $\alpha^B$ than $P^B_{max}$, since PAR in the Labrador Sea is rarely sustained at high enough intensity for the photosynthetic rate to reach saturation at $P^B_{max}$.
In contrast, although the VGPM model is mostly driven by Chl-a, its sensitivity to SST surpassed that of Chl-a from mid July to September. Unlike other parameters, the model sensitivity to SST exhibited a pronounced seasonal pattern, whereas the sensitivity trend of other parameters remained relatively random and flat over time. Based on the model configuration, figure 7 in (Behrenfeld & Falkowski, 1997) showed that $P^B_{opt}$ shifted only ~ 2 mg C (mg chl-a)$^{-1}$ h$^{-1}$ over the range of SST between 0 and 8℃, while the change was more dramatic for water above 8℃. This pattern coincided with the sudden increase of $|S_{VGPM}|$ after SST increased past 8℃ (Figure 6e), suggesting that the seasonal pattern is related to the nonlinear temperature dependence of $P^B_{opt}$ in the model. Therefore, the VGPM model revealed distinct photosynthetic responses compared with the Webb and the BIO models.

## 5.2. What causes the NPP difference?

The magnitude of differences between depth-integrated $NPP_{Webb}$ and $NPP_{BIO}$ was mainly attributed to the methods used to estimate P-I parameters within the timeframe of this study. In the BIO model, P-I parameters are selected based on oceanographic consideration revealed by both satellite-derived Chl-a concentration and sea surface temperature, while the Webb model used a monthly climatology based on a large archive (128 data points) in the Labrador Sea. This led to large differences of P-I parameters (Figure 4b & 4c) and created a considerable gap between $NPP_{Webb}$ and $NPP_{BIO}$ that is strongly correlated to $\Delta\alpha^B$ (Figure 7b). Comparatively, the impact of $\Delta P^B_{max}$ was not as important as $\Delta\alpha^B$ on NPP due to limited available light for photosynthesis in high latitude regions. The agreement between depth-integrated $NPP_{Webb}$ and $NPP_{BIO}$ improved significantly when the same set of P-I parameters were assigned to both models (Figure 8b). This demonstrates that the difference of *in situ* and satellite-based NPPs could be minimized with a better characterization of P-I parameters in the region, especially for periods outside the bloom season when the data are scarce.

When applied the same set of P-I parameters in the BIO and Webb model, time became the most influential factor in $\Delta NPP_{BIO\text{-}Webb^*}$. While Chl-a and PAR were ecologically relevant, their lower importance score indicated that time captured much of the variability observed in $\Delta Chl_{POLY4\text{-}prof}$ and $\Delta PAR$. The temporal variability was linked to limited satellite detection in the later season, notably around mid September, when PAR began to drop below 20 µmol photons m$^{-2}$ d$^{-1}$. This was supported by the outliers in October observed from the comparison of Chl-a (Figure 3d) and NPP (Figure 8b), as well as the heightened sensitivity of NPP to the change of satellite PAR in the BIO and VGPM model around the same time (Figure 6). Although satellite data were gap-filled using the DINEOF method, persistently lower data coverage, such as in later months or more northerly areas, resulted in higher estimate errors.

The discrepancy between $NPP_{VGPM}$ and $NPP_{Webb}$, however, told a different story. A lack of consistency in temporal patterns was observed in $\Delta NPP_{VGPM\text{-}Webb}$ for both surface and depth-integrated time series (Figure 5c & 5d). This is due to increasing bias in the SST-dependent $P^B_{opt}$, and the absence of the $\alpha^B$ term in the VGPM model (Figure 7a). The empirical relationship between SST and $P^B_{opt}$ has been considered the Achilles' heel of the VGPM model (Hernández-Hernández et al., 2025; Z. Lee & Marra, 2022). Numerous studies have reported poor agreement between *in situ* and satellite $P^B_{opt}$ (Hernández-Hernández et al., 2025; Regaudie-de-Gioux et al., 2019), suggesting the sole dependency of $P^B_{opt}$ on SST oversimplified the complexity within $P^B_{opt}$. Since the VGPM model itself is quite sensitive to SST, this led to the findings of the unexplained pattern in $NPP_{VGPM}$ during non-bloom periods when SST rarely varied. As a result, $P^B_{opt}$ might not be well-suited for larger-scale application at high-latitude regions, including the Labrador Sea, as the photosynthetic response was biased towards regions with warmer water. Moreover, the dominance of $\alpha^B$ as a driver of $\Delta NPP_{VGPM\text{-}Webb}$ highlighted its importance of accurately representing photosynthetic efficiency under lower PAR conditions, particularly in high-latitude regions during the winter and early spring. This suggests that incorporating $\alpha^B$ may be more important than relying primarily on the optimal growth rate as portrayed by the VGPM model.

## 5.3. Why VGPM and BIO produce different NPP estimates?

The major difference between the VGPM and BIO models is primarily in the representation of photosynthetic terms. The VGPM model used the photosynthetic rate $P^B_{opt}$ derived from a global empirical relationship with SST (Eppley, 1972), which does not account for other key variables that could affect photosynthetic efficiency (e.g., Chl-a, PAR, nutrients). The BIO model adopted per-pixel assignment of $\alpha^B$ and $P^B_{max}$ using SST and Chl-a as predictors based on a clustering analysis of phytoplankton community composition and trophic levels (Devred et al., 2025). Moreover, the VGPM model did not include $\alpha^B$, a key parameter to describe photosynthesis at low light conditions. The BIO model not only included $\alpha^B$ but adjusts for spectral dependence by assuming its spectral shape resembles the shape of the phytoplankton absorption spectrum (Eq. 6). These configuration differences explained why the VGPM model was most responsive to SST, whereas the BIO model displayed stronger sensitivity to $\alpha^B$.

Although the VGPM and BIO models utilized raw data from the MODIS-Aqua satellite, they applied different Chl-a retrieval algorithms (OCI and POLY4). The regionally tuned $Chl_{POLY4}$ in the BIO model better reproduced the June-July bloom than global $Chl_{OCI}$ in the VGPM model (Figure 2c). Since the relative error of NPP is highly correlated to the errors of surface Chl-a concentration (Campbell et al., 2002), regionally tuned products can reduce the error propagation down to the NPP estimates.

More importantly, both models differed in how surface observations propagate throughout the euphotic zone. The VGPM model extrapolates all variables from the surface to depth by assuming they are constant within the euphotic zone. While the BIO model maintains the assumption of constant Chl-a concentration and photosynthetic parameters throughout the water column, it resolved the subsurface light field using a radiative transfer model (Gregg & Carder, 1990) by modelling light attenuation in the water column as a function of absorption and scattering by water, phytoplankton, colored dissolved organic matter (yellow substances), and detritus (Devred et al., 2025). As a result, the BIO model exhibited 6 times higher sensitivity to PAR than the VPGM model (Figure 6c, f), indicating that the BIO model captured the role of light availability in regulating NPP more realistically.

## 5.4 Which is more important: good model or good data?

Our study demonstrated that the impact of an input variable on NPP is determined by its magnitude, its representation within the model and the model's sensitivity to changes in that variable. Consequently, mismatches among NPP estimates may arise not only from the differences in the magnitude of input variables, but from the extent to which these differences propagate through the models. The comparison between Webb and BIO models provided a clear example. The P–I parameters assigned by the two models differed substantially over the same periods, while both models also showed strong sensitivity to $\alpha^B$, resulting in the differences of $\alpha^B$ contributed the most to the divergence between $NPP_{Webb}$ and $NPP_{BIO}$. When both models are applied with matching P-I parameters, their NPP estimates become more consistent with each other, supporting the discrepancy of P-I parameters passed down to the difference of their NPP estimates. In this case, improving the agreement of their P-I parameters has greater impacts than other input variables.

Unlike the BIO model, the VGPM model derives the photosynthetic response through the SST-dependent $P^B_{opt}$ term rather than $\alpha^B$. Although the VGPM model showed high sensitivity to $P^B_{opt}$, but the influence of $\alpha^B$ from the Webb model was comparatively stronger on $\Delta NPP_{VGPM-Webb}$ (Figure 7a). As a result, the discrepancy of $NPP_{VGPM}$ and $NPP_{Webb}$ can not be solely explained by the differences in the magnitude of input variables. Instead, it reflects the impact of model parameterization on photosynthetic rates that would be appropriate for the region, which $\alpha^B$ would matter more than $P^B_{opt}$ in this case. The contradictory responses of NPP differences from both model comparisons further demonstrate that similar discrepancies in NPP estimates might not come from the same source of bias.

Our findings suggest that neither “good data” or “good model” alone is sufficient to be an independent solution. Regional observations can provide more representative inputs for NPP models; however, their benefits depend on the ability of the models to translate those inputs and the associated regional processes into realistic NPP estimates. Hence, reconciling in situ and satellite NPP estimates requires considering both the regional validity of model inputs and the model-specific responses of NPP to those inputs, especially when the model sensitivity varies seasonally or nonlinearly. Priority should be given to those inputs that differ significantly with regional observations and exert a strong influence on the NPP estimates. However, these conclusions are based on chlorophyll-based models evaluated at a single site in the Labrador Sea, they should be tested in other regions, over different time periods, and using other NPP models.

## 5.5 Limitations and Recommendations

Overall, this study was limited to chlorophyll-based primary productivity models. The goals of this study were to emphasize the importance of using regional data sources for NPP models and to assess the applicability of model outputs within the Labrador Sea. By examining the strength and weakness of each model, this study aimed to bring forward the best option available for the subpolar Atlantic region. We recognized the limitation of using monthly climatology of P-I parameters in the estimates of $NPP_{Webb}$, as the 40-years dataset were unequally distributed over the seasons due to ship time restriction. In fact, most measurements were made in spring (May - June), resulting in a bias of seasonal distribution in our dataset. Improved seasonal coverage of ship-based P-I parameters (Brewin et al., 2023), particularly outside the bloom seasons, is therefore necessary to improve model robustness. Similarly, the BIO model is regionally tuned to the Labrador Sea and more challenging to apply at synoptic scales, given the spatio-temporal variation in photosynthetic parameters. However, new approaches based on machine learning may solve the issue (Britten et al., 2025).

In the past 20 years, newer productivity models, such as carbon-based (CbPM) and absorption-based (AbPM) models, were developed as an improvement upon the VGPM model (Silsbe et al., 2016; Westberry et al., 2008). Most productivity models started to use globally based empirical relationships to estimate NPP that may be sufficient for global studies, but may not necessarily be suitable for regional-scale studies. Therefore, applying global relationships to biogeochemically heterogeneous regions without adequate validation data might lead to poorly established proxies and inaccurate NPP estimates. An important lesson from this study is that reliable regional NPP estimates require both well-characterised regional parameters and a model that appropriately translates those parameters into NPP. For the Labrador Sea, the immediate priorities are to improve regional and seasonal characterization of P-I parameters, evaluate the suitability of SST-dependent $P^B_{opt}$ in the region and consider includ depth-resolved Chl-a profiles in satellite NPP estimates. Particularly, the limited availability of P-I measurements outside spring bloom restricts the characterization of seasonal variability of phytoplankton physiology. Carr et al. [57] emphasized the need for improved parameterization of $P^B_{max}$ and a deeper understanding of the influence of sea surface temperature (SST) on photosynthesis, suggesting a potential update to the relationship between SST and $P^B_{opt}$, particularly to better represent high-latitude regions.

Future research should prioritize observation strategies that improve regional and seasonal observations in the NPP models. Long-term, high-frequency and depth-resolved observations from fixed moorings can provide continuous characterization of seasonal and short-term variability from bio-optical and biogeochemical properties, including spectral irradiance, phytoplankton absorption, dissolved oxygen and carbonate variables, such as total alkalinity, pH and $pCO_2$. Mobile platforms, such as biogeochemical Argo floats, can extend these observations across broader spatial gradients. Targeted ship-based sampling should remain essential to provide direct P-I measurements. Together, these complementary approaches could fill temporal, spatial, and vertical gaps that currently limit the development and evaluation of regional NPP models. These observations can also support hybrid modelling approaches that integrate multiple data sources and environmental variables to estimate P-I parameters where direct measurements are sparse. Incorporating phytoplankton community structure and a range of environmental factors, such as SST, light, and nutrients, into model frameworks can provide a more ecologically representative basis for assigning P–I parameters across diverse marine ecosystems (Richardson et al., 2016).

# 6. Conclusion

This study presents a comparative analysis of net primary productivity (NPP) time series derived from two satellite-based models and an in situ-based model in the Central Labrador Sea between May and November 2016. A considerable difference in NPP magnitude is observed between satellite- and in situ-based time series, which is strongly associated with differences in the physiological terms ($P^B_{opt}$, $P^B_{max}$, $\alpha^B$) and model sensitivity to these parameters. Assigning a common set of P-I parameters improved the agreement between $NPP_{Webb}$ and $NPP_{BIO}$, demonstrating the critical importance of consistent and regionally model parameterization. Future improvements should prioritize regional characterization of model inputs that strongly drive the model on NPP estimates. For Labrador Sea, improved characterisation of seasonal variability in $\alpha^B$ across seasons should be prioritised. However, the limited availability of P-I measurements outside the spring bloom season restricts our ability to reliably estimate seasonal NPP and, by extension, the strength of regional BCP.

Addressing these gaps requires observation strategies designed to constrain the sources of uncertainty in NPP estimates. Fixed moorings can provide long-term, high frequency and depth-resolved observations, while mobile platforms can extend the spatial coverage. Increasing ship-based measurements across different seasons is essential to capture the temporal variability of the P-I parameters. These observations can further support the development of modeling techniques that integrate environmental drivers, such as sea surface temperature (SST), Chl-a, nutrient concentrations and phytoplankton community structure, to predict P-I parameters where data coverage is currently limited.

Since NPP defines the production of organic carbon that can contribute to export, uncertainties in NPP estimates can propagate into the estimates of carbon export and may introduce systematic biases. Given the Labrador Sea's role as a major $CO_2$ sink, this study highlights the need to better constrain the variability of phytoplankton physiology to improve regional carbon budget estimates. Stronger emphasis on regional parameter characterization will improve estimates of NPP, and strengthen the assessment its role in carbon export, ecosystem dynamics, and ocean biogeochemical cycles.

# Funding Source

This work was supported by the Ocean Frontier Institute (OFI) through "Northwest Atlantic Biological Carbon Pump" (NWABCP) to Zoe Finkel (Dalhousie), NSERC through the Discovery Grant to Douglas Wallace and the Advancing Climate Change Science in Canada (ACCSC) project: "Quantifying and Predicting Canada's Marine Carbon Sink" to Roberta Hamme (Univ. Victoria). The SeaCycler platform was developed with funding from the US NSF OCE Technology grant OCE0501783 to Uwe Send (UCSD).

# Acknowledgement

We thank Michael Dowd (Dalhousie) for support in statistical analysis, and the contribution of the Bedford Institute of Oceanography (Fisheries and Oceans Canada).

# Data statement

Data and code will be made available on a public repository.

# Glossary

| Variables | Definition |
|---|---|
| $Chl_{OCI}$, $Chl_{POLY4}$ | Satellite chl-a estimated by OCI and POLY4 ocean colour algorithms (mg $m^{-3}$) |
| $Chl_{prof}$ | Depth-resolved chl-a observed by SeaCycler |
| $E_d(\lambda)$ | Downwelling irradiance measured at $\lambda$ |
| $K_d(\lambda)$ | Diffuse attenuation coefficient of downwelling irradiance at $\lambda$ ($m^{-1}$) |
| $NPP_{webb}$ | Net Primary Production from the Webb model (equation 2) |
| $NPP_{VGPM}$ | Net Primary Production from the VGPM model (equation 3) |
| $NPP_{BIO}$ | Net Primary Production from the BIO model (equation 5) |
| PAR | Photosynthetic Active Radiation (µmol photon $m^{-2}$ $d^{-1}$) |
| $P^B_{opt}$ | Optimal photosynthetic capacity |
| $P^B_{max}$ | Maximum photosynthetic rate normalized by chl-a concentration |
| SST | Sea Surface Temperature |
| $\alpha^B$ | Photosynthetic efficiency under minimal light levels |
| $\lambda$ | Wavelength (nm) |

# Supplementary Materials

**Table S1.** A list of sensors equipped on the SeaCycler mooring during 2016 deployment.

| Types | Variables | Sensors |
|---|---|---|
| Bio-optical | Chlorophyll a (chl-a) concentration, particulate backscattering ($b_{bp}$), CDOM | WetLab ECO-FLBBCD sensor |
| | Monochromatic downwelling irradiance at 490nm ($E_d$490) | Satlantic PAR sensor (modified for 490 nm) |
| | Transmittance, particulate scattering ($c_p$) | WetLab C-Star transmissometer |
| Chemical | Dissolved oxygen | 1. SeaBird dissolved oxygen sensors (SBE 43, 63)<br>2. Aanderaa oxygen optode |
| | Nitrate | Deep SUNA ocean nitrate sensor |
| | Dissolved Inorganic Carbon | 1. ProOceanus ProCV $CO_2$ sensor<br>2. Aanderaa p$CO_2$ optode |
| | Salinity (conductivity) | SBE 19plus V2 SeaCAT |
| Physical | Temperature, Depth | SBE 19plus V2 SeaCAT |

**Table S2.** Summary of NPP inputs used in this study

| NPP output | $NPP_{Webb}$ | $NPP_{VGPM}$ | $NPP_{BIO}$ |
|---|---|---|---|
| **Chl a** | **$Chl_{prof}$**<br><br>In situ chl-a profiles | **$Chl_{OCI}$**<br><br>MODIS-aqua R2022 version of chl-a product processed by the OCI algorithm | **$Chl_{POLY4}$**<br><br>MODIS-aqua satellite chl-a processed by OCx algorithm, with coefficients tuned by "POLY4" relationship for NWA (Clay et al., 2019).<br><br>Observational gaps are |
| | | | filled by the DINEOF method (Alvera-Azcárate et al., 2011) |
| **PAR** | In situ $E_d(490)$ profiles converted to PAR profiles using the COART model (Jin et al., 2006) | Satellite PAR | Modeled subsurface light field based on Gregg & Carder (1990) and scaled up to measured satellite PAR |
| **Photosynthetic Parameters** ($P^B_{max}$, $\alpha^B$, $P^B_{opt}$) | Ship-based incubation data from AZOMP time series between 1977 and 2019.<br>- P-I curves are estimated using piCurve package (Amirian & Irwin, 2025)<br>- $P^B_{max}$ and $\alpha^B$ are spatially averaged within a 2 degree bin surrounding the location of the SeaCycler mooring | $P^B_{opt}$ is computed as a function of satellite SST | Ship-based incubation data from AZOMP time series between 2014 and 2022.<br>- P-I curves are estimated using piCurve package (Amirian & Irwin, 2025)<br>- $P^B_{max}$ and $\alpha^B$ are selected based on the criteria of characterized biomes (Devred et al., 2025) |
| $Z_{eu}$ | Derived from in situ profiles of PAR using 1% PAR criterion | Chl-based empirical relationship | Defined as 1% surface PAR, with light attenuation at each depth modeled as a function of absorption and scattering in the water column |

**Fig S1.** Monthly, full depth averages of P-I parameters for Pmax (left), alpha (middle) and $I_k$ (right) for in situ NPP. Monthly averages of each parameter are marked with a cross (X) while the annual averages are marked as dashed lines.

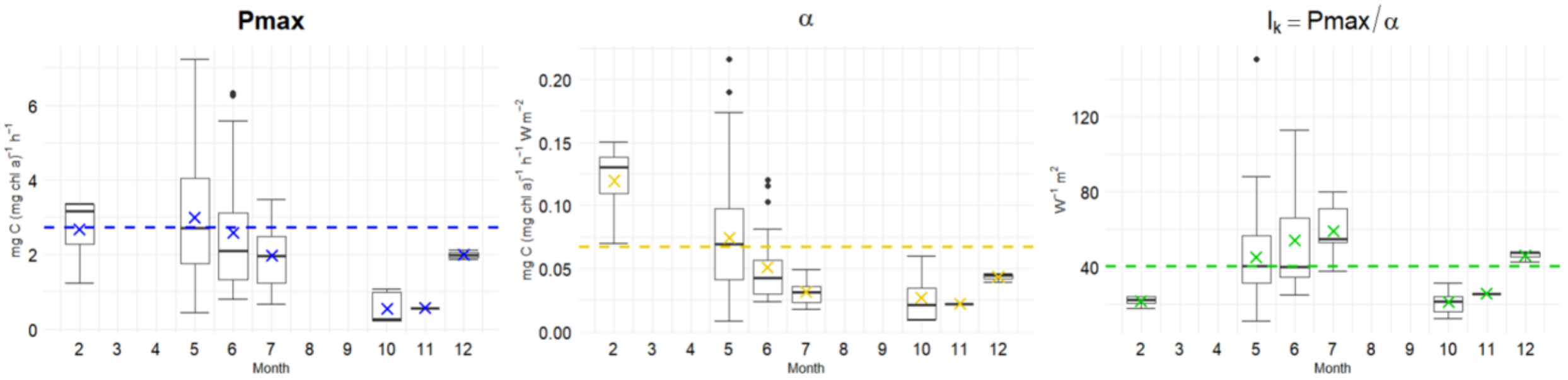

**Fig S2.** Monthly, 10m binned (for top 30m) averages of P-I parameters for $P^B_{max}$ (left), $\alpha^B$ (middle) and $I_k$ (right). Monthly averages of each parameter are marked with a cross (X) while the annual averages are marked as dashed lines.

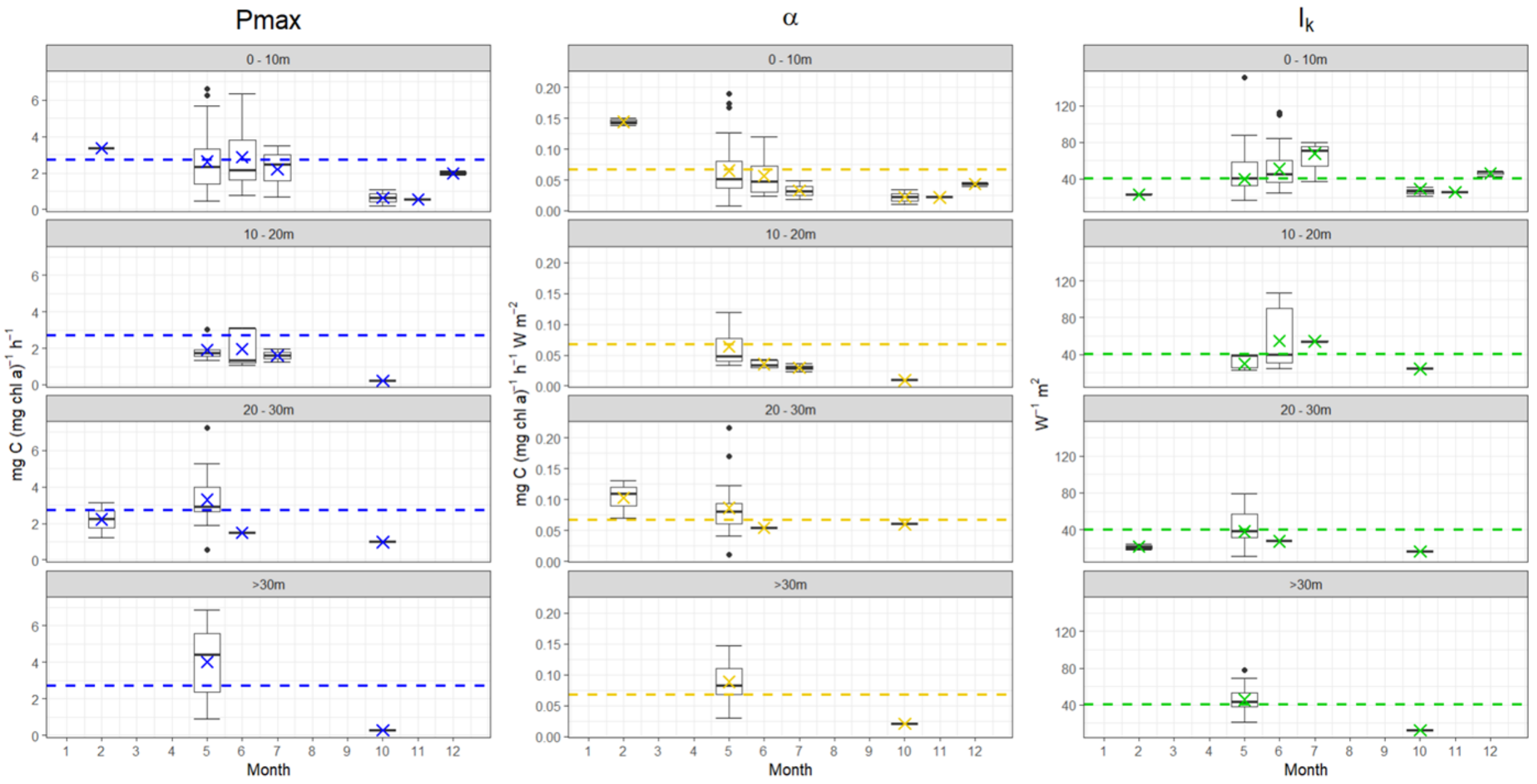

**Fig S3.** (a) Time series of $NPP_{Webb}$ using three different methods to incorporate P-I parameters: (1) annual, full depth averages (red), (2) monthly, full depth averages (blue, Fig.S1), and (3) monthly, 10m binned averages (green, Fig.S2). (b) Difference of NPP (in %) between annual and monthly averages of P-I parameters, showing the impact of temporal resolution of P-I parameters on NPP. (c) Difference of NPP (in %) between full depth and 10m binned averages of P-I parameters.

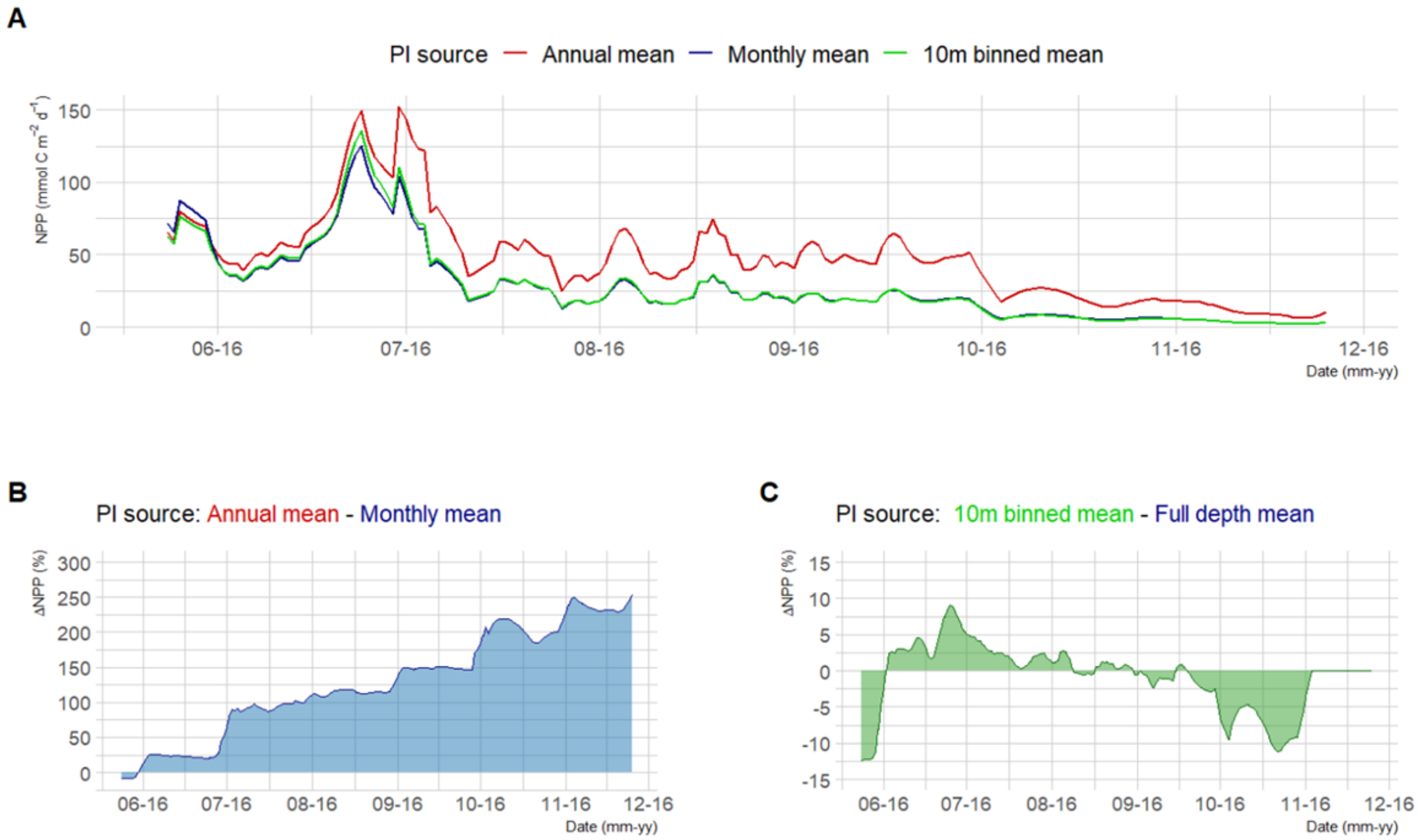

**Figure S4.** Linear correlation between original ($NPP_{BIO}$) and updated depth-integrated NPP from the BIO model using in situ $Chl_{prof}$ from SeaCycler ($NPP_{BIO,Chl\ prof}$). The scatter points are colour coded by the depth of seasonal chl-a maxima observed from in situ $Chl_{prof}$.

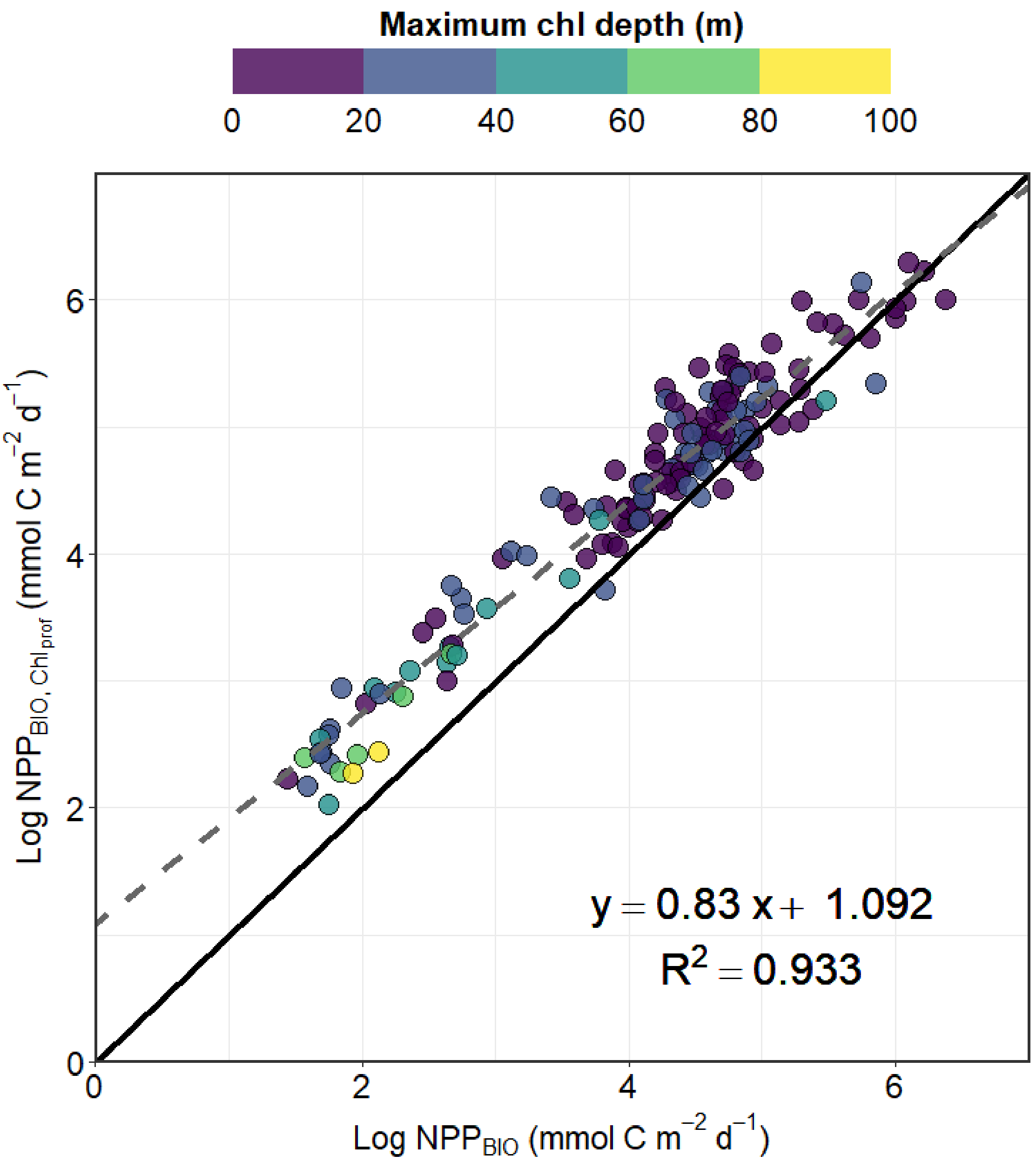